\documentclass[12pt,preprint,letterpaper]{aastex}
 
\shorttitle{Magnetized Bondi Accretion}
\shortauthors{Cunningham, et al.}

\def\solar{\ifmmode _{\mathord\odot}\else $_{\mathord\odot}$\fi}
\def\msun{\ifmmode {\rm M}_{\mathord\odot}\else $M_{\mathord\odot}$\fi}
\def\lsun{\ifmmode {\rm L}_{\mathord\odot}\else $L_{\mathord\odot}$\fi}
\def\sol{\ifmmode {\mathord\odot}\else ${\mathord\odot}$\fi}

\renewcommand{\vec}[1]{{\bf #1}}

\newcommand{\beq}	{\begin{equation}}
\newcommand{\eeq}	{\end{equation}}
\newcommand{\beqa}{\begin{eqnarray}}
\newcommand{\eeqa}{\end{eqnarray}}
 
\newcommand{\dis}{\displaystyle}
\newcommand{\e}	{$^{-1}$}

\def\simlt{\lower.5ex\hbox{$\; \buildrel < \over \sim \;$}}
\def\simgt{\lower.5ex\hbox{$\; \buildrel > \over \sim \;$}}
\def\la{\simlt}
\def\ga{\simgt}

\def\vecnabla{
              \setbox1=\hbox{$\bigtriangledown$}
                           \raise.45ex\hbox{$\bigtriangledown$\hskip-.97\wd1
                           $\bigtriangledown$\hskip-.97\wd1
                           $\bigtriangledown$\hskip-.97\wd1}
                           \raise.47ex\hbox{$\bigtriangledown$}}
\def\grad{\vecnabla}

\newcommand{\vecB}	{{\bf B}}

\newcommand{\ppbyp}[2]	{{{\partial#1}\over{\partial#2}}}
\newcommand{\alfven}    {{Alfv$\acute{\rm e}$n}}
\newcommand{\bro}		{B_{r0}}
\newcommand{\bto}		{B_{\theta 0}}
\newcommand{\esc}	   {{\rm esc}}

\newcommand{\rb}		{{r_{\rm B}}}
\newcommand{\rab}		{r_{\rm AB}}
\newcommand{\roa}	   {r_{0a}}
\newcommand{\rpo}	   {r_{\Phi,\,1}}
\newcommand{\rpt}	   	   {r_{\Phi,\, 2}}
\newcommand{\tb}		{{t_{\rm B}}}
\newcommand{\va}		{v_{\rm A}}
\newcommand{\vk}		{v_{\rm K}}
\newcommand{\seq}	   {\Sigma_{\rm eq}}
\newcommand{\varcr}	   {\varpi_{\rm cr}}
\newcommand{\varocr}	   {\varpi_{\infty,\,\rm cr}}
\newcommand{\llb}		  {\lambda_{\rm low\;\beta}}

\begin{document}
\title{Radiatively Efficient Magnetized Bondi Accretion}

\author{Andrew J. Cunningham\altaffilmark{1}, Christopher F. McKee\altaffilmark{2,3}, Richard I. Klein\altaffilmark{1,2}, Mark R. Krumholz\altaffilmark{4}, Romain Teyssier\altaffilmark{5}}

\altaffiltext{1}{Lawrence Livermore National Laboratory, Livermore, CA 94550}
\altaffiltext{2}{Department of Astronomy, University of California Berkeley, Berkeley, CA 94720}
\altaffiltext{3}{Department of Physics, University of California Berkeley, Berkeley, CA 94720}
\altaffiltext{4}{Department of Astronomy and Astrophysics, University of California Santa Cruz, Santa Cruz, CA 94560}
\altaffiltext{5}{Service d'Astrophysique, CEA Saclay, 91191 Gif-sur-Yvette, France}

\email{ajcunn@gmail.com}

\begin{abstract}
We have carried out a numerical study of the effect of large scale
magnetic fields on the rate of accretion from a uniform, isothermal
gas onto a resistive, stationary point mass.  Only mass, not magnetic
flux, accretes onto the point mass. The simulations for this study
avoid complications arising from boundary conditions by keeping the
boundaries far from the accreting object.  Our simulations leverage
adaptive refinement methodology to attain high spatial fidelity close
to the accreting object.  Our results are particularly relevant to the
problem of star formation from a magnetized molecular cloud in which
thermal energy is radiated away on time scales much shorter than the
dynamical time scale.  Contrary to the adiabatic case, our simulations
show convergence toward a finite accretion rate in the limit in which
the radius of the accreting object vanishes, regardless of magnetic
field strength.  For very weak magnetic fields, the accretion rate
first approaches the Bondi value and then drops by a factor $\sim 2$
as magnetic flux builds up near the point mass. For strong magnetic
fields, the steady-state accretion rate is reduced by a factor $\sim
0.2 \beta^{1/2}$ compared to the Bondi value, where $\beta$ is the
ratio of the gas pressure to the magnetic pressure.  We give a simple
expression for the accretion rate as a function of the magnetic field
strength. Approximate analytic results are given in the Appendixes for
both time-dependent accretion in the limit of weak magnetic fields and
steady-state accretion for the case of strong magnetic fields.
\end{abstract}


\section{Introduction} \label{introduction}
Accretion of a background gas onto a central gravitating body is of
central importance in astrophysics.  Examples range from protostellar
accretion from molecular cores to accretion of interstellar gas in
galactic nuclei.  The classical late-time solution for the case of a
central point of mass $M_*$ immersed in a uniform, initially static,
unmagnetized gas was given by \citet{bondi} as
\begin{eqnarray}
\dot{M}_{\rm B} &=& 4 \pi \lambda \rb^2 \rho_\infty c_\infty \\
\rb &=& \frac{GM_*}{c^2_\infty}
\end{eqnarray}
where $c_\infty$ and $\rho_\infty$ are the sound speed and density of
the background gas, $\dot{M}_{\rm B}$ is the steady-state rate of
accretion onto the central particle, $\rb$ is the Bondi length which
characterizes the dynamical length of the inflow and $\lambda$ is a
dimensionless parameter that depends on the equation of state of the
background gas.  For the isothermal case, $\lambda =
\textrm{exp}(1.5)/4$.  The Bondi time $\tb = \rb/c_\infty$ defines the
dynamical time for this accretion process.  This basic model has been
extended to more general cases by numerous authors.  These
generalizations include non-stationary central particles
\citep{bondihoyle,shima,ruf1,ruf2}, the cases of ambient gas with net
vorticity \citep{krumholzvorticity}, turbulent ambient gas
\citep{krumholzturbulent}, magnetized accretion from ambient gas
threaded by both large \citep{igu,pang} and small
\citep{shapiro,igusmall} scale magnetic field topologies, the case of
a turbulent, magnetized ambient gas \citep{shch}, and the case of
accretion onto magnetized stars
\citep{toropin,Ustyugova06,Kulkarni08,Romanova08,Long11,Romanova11},
to name a few.

Stars form via gravitational collapse, at least initially
\citep{mckeeostriker}.  Thereafter, gas may accrete onto the star from
the ambient medium.  If the star has a supersonic motion relative to the
ambient medium, this subsequent accretion is negligible \citep{kmk05},
but if the star is moving slowly, the accretion can be significant,
which forms the basis for the competitive accretion model for star
formation (e.g., \citealp{bon97}).  There exists ample evidence that
the gas in molecular clouds and cores is threaded by strong magnetic
fields \citep{crutcher,mckeeostriker}.  Furthermore, star forming
molecular clouds are well characterized as ``radiatively efficient''
in that gas heating due to compressional motion is rapidly radiated by
thermally excited dust and molecules.  These considerations thus
motivate the study of Bondi-type accretion of an isothermal gas
threaded by an initially uniform magnetic field onto a point mass.  We
address this problem with the {\ttfamily RAMSES} magnetohydrodynamic (MHD) code
\citep{ramses} and conduct a parameter study over a range of magnetic
field strengths thought to be relevant to star formation.  Our
simulations leverage the adaptive mesh refinement (AMR) capability of
the code to retain high spatial resolution close to the accreting
object while keeping the boundaries of the computational domain far
from the accreting object.  We discuss the results of mesh convergence
studies and compare our numerical results against analytic
calculations in the limiting case of a dynamically weak magnetic field
to verify our calculations.  We also compare our numerical results
against simple analytic approximations in the case of a strong
magnetic field.

\section{Numerical Setup} \label{setup}
Our numerical models consist of a Cartesian computational domain that
extends from $-25 \rb~\textrm{to}~25 \rb$ in each direction.  The
domain is initialized with an isothermal, perfectly conducting,
uniform collisional gas with initial magnetic field in the $\hat{z}$
direction.  We consider the cases with an initial thermal to magnetic
pressure ratio, $\beta=8 \pi P_o / \vec{B}^2$, of 1000, 100, 10, 1,
0.1 and 0.01.  The {\ttfamily RAMSES} code has been used to evolve this
state forward according to the equations of ideal, isothermal MHD,
\begin{eqnarray}
\frac{\partial \rho}{\partial t} + \grad \cdot \rho \vec{v} &=& -S_M \\
\frac{\partial \rho \vec{v}}{\partial t} + \grad \cdot (\rho \vec{v}\vec{v}) + \grad \left(P + \frac{\vec{B}^2}{8 \pi} \right) - \frac{(\vec{B} \cdot \grad) \vec{B}}{4 \pi} &=& -\frac{G M_* \rho~ \hat{\vec{x}}}{\vec{x}^2} \\
\frac{\partial B}{\partial t} - \grad \times (\vec{v} \times \vec{B}) &=& 0 \\
P &=& \rho c^2,
\end{eqnarray}
where $\rho$ is the gas density, $\vec{v}$ is the velocity, $\vec{B}$
is the magnetic field, $P$ is the thermal pressure and $c$ is the
isothermal sound speed.  These equations include the gravitational force
due to a point particle of mass $M_*$ of $\vec{F}_g = -G M_* \rho~
\hat{\vec{x}}/\vec{x}^2$.

The key assumption we make in our treatment is that the point mass
accretes mass, but not flux. Observations show that the magnetic flux
in young stellar objects is orders of magnitude less than that in the
gas that formed these objects, implying that flux accretion is very
inefficient, presumably due to non-ideal MHD effects, including
reconnection \citep{mckeeostriker}.  We model mass accretion onto the
central point mass by including a mass sink term but no flux sink term
inside a radius, $r_{\rm acc} =4\Delta x$, equal to four grid zones on
the finest AMR level.  The effect of the accreting particle is coupled
to the dynamical equations through the source term,
\begin{equation}
  S_M = \left\{\begin{array}{ll}
  \dis\frac{1}{\Delta t} \,\textrm{max}\left(\rho - \frac{\vec{B}^2}{4 \pi v_{\rm A,max}^2}\, ,\, 0\right)& \textrm{if ~} |\vec{x}| < r_{\rm acc} \\
  0 & \textrm{otherwise,}
  \end{array} \right. \label{sm}
\end{equation}
where $\Delta t$ is the time step on the finest AMR grid level and
$v_{\rm A,max}$ is the maximum \alfven\ speed, $B/(4\pi\rho)^{1/2}$,
within a radius of $6 \Delta x$ around the accreting particle.  Under
this construction, the accreting particle absorbs all but enough of
the mass entering the accreting particle radius so that the local
\alfven\ speed never exceeds $v_{\rm A,max}$.  Thus, the accreting
particle always absorbs the largest quantity of mass in the local
region possible without introducing new local extrema in the \alfven\
speed.  This prevents the accretion source from imposing a stringent
(or vanishingly small) constraint on the maximum numerically stable
time-step at the expense of some artificial clipping of the \alfven\
speed in the inner few zones around the accreting particle.  We note
that in all of the models considered in this paper, the initial gas
density is sufficiently low that the total mass accreted onto the
central particle is negligible compared to $M_*$ and that self-gravity
in the ambient medium may be neglected.

We discretize the numerical domain onto a base level grid of $64^3$.
For the purposes of describing the initial mesh we will denote this
level as $l=0$.  We note, however, that the {\ttfamily RAMSES} AMR
implementation uses an oct-tree data structure for level traversals
that always denotes level indices by the base 2 logarithm of their
resolution.  In our models, $l_{\rm RAMSES} = \textrm{log}_2 64 + l =
6 + l$.  Successive levels are chosen for refinement by an increment
of $2^3$ in grid zone density in a geometrically nested fashion
according to the criterion
\begin{equation}
r_l < \frac{25 \rb}{2^{l}},
\end{equation}
where $r_l$ indicates the radius of the spherical refined region on
the level $l$.  We further impose the additional criterion that any
zones containing steep density gradients $\vec{\grad} \rho \cdot
\vec{\Delta x}/\rho > 1/2$ are also refined, independent of location.
This second refinement criterion is met only at late times after
non-axisymmetric flow patterns have set in, and it triggers only on
transient flow features.  Most of the models were refined to a maximum
level $l=8$ for an effective resolution of $64\times 2^8/50\simeq 328$
zones per thermal Bondi radius on the finest level.  In the cases with
a strong initial magnetic field, it is also useful to consider the
numerical resolution on the scale of the ``\alfven-Bondi'' radius,
\begin{equation}
\rab = \frac{GM_*}{\va^2} =\frac 12 \beta \rb.
\end{equation} 
The finest mesh resolution per Bondi radius, mesh resolution per
\alfven-Bondi\ radius and total simulated time for each model is
tabulated in Table \ref{t1}.  We note that the magnetic length scales
are well resolved for all but the case of $\beta=0.01$.  We therefore
will consider only the models with $\beta \ge 0.1$ for the majority of
the analysis presented in this paper.  The $\beta = 0.01$ model is
used only to extract an estimate of the steady accretion rate over a
wider range of magnetic field strengths.  We do note, however, that
numerical mesh convergence studies have shown our models to be within
the range of asymptotic convergence with a Richardson-extrapolation
error estimate on the average accretion rate of $14\%$
or less at late times.  A detailed discussion of the numerical convergence
properties of our models is presented in Appendix \ref{convergence}.
Each of the models were run to a final time $t_{\rm end}$ sufficiently
long to attain a statistically steady accretion rate onto the central
particle.

\begin{deluxetable}{l l l l}
  \tablewidth{0pt}
  \tablecaption{Simulation Parameters.\label{t1}}
  \startdata
  \tableline
  $\beta$ & $r_{\rm B}/\Delta x$ & $\rab/\Delta x$ & $t_{\rm end}/\tb$\\
  \tableline
  $\infty$ (hydro) & 328 & N/A & 3 \\
  1000  & 82 & 41000 & 22 \\
  100  & 82  & 4100 & 15 \\
  10   & 328 & 1640 & 3 \\
  1    & 328 & 164 & 3 \\
  0.1  & 328 & 16.4 & 3 \\
  0.01 & 328 & 1.64 & 1.5 \\
  \enddata
  \tablewidth{\textwidth}
\end{deluxetable}

\section{Results}
\subsection{Morphology} \label{morphology}
\begin{figure}[tpb]
\begin{center}
\includegraphics[clip=true,width=0.75\textwidth]{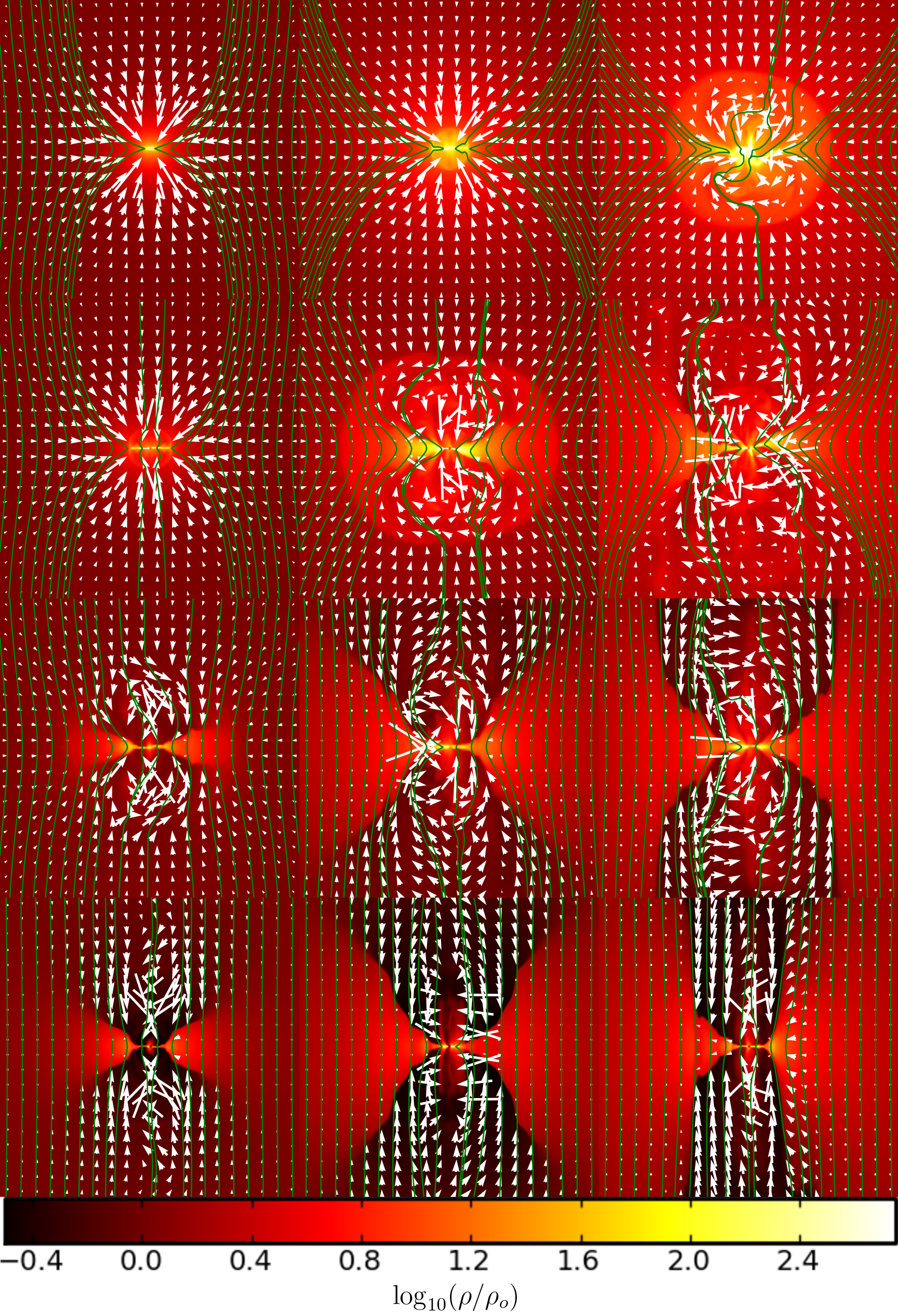}
\end{center}
\caption{Slices in the y-z plane showing the inner $(2 \rb)^2$ of the
  numerical models with initial magnetic field strengths of $\beta=100$,
  $\beta=10$, $\beta=1$, and $\beta=0.1$ from top to bottom.  The
  state of the numerical models are shown at the times $t =
  0.5,~1.5,~\textrm{and}~3.0~\tb$ from left to right.  The colormap
  indicates $\log_{10}(\rho / \rho_o)$, green lines represent magnetic
  flux tubes drawn from equidistant foot-points in the midplane and the
  white arrows indicate the flow pattern in the plane of the
  slice.  \label{f1}}
\end{figure}

We begin by discussing the gross morphological flow features and their
development for each of the numerical models.  These flows are well
illustrated by slices in the y-z plane of density, the direction of
magnetic flux and velocity as shown at several times for each model in
Figure \ref{f1}.  Initially parallel magnetic fields are amplified as
they are dragged inward by the global accretion flow, eventually
suppressing accretion in the equatorial plane.  Inflow along magnetic
field lines, on the other hand, is uninhibited by magnetic pressure.
This flow configuration leads to the evacuation of gas from the
poleward directions into a thin, dense, irrotational disk in the
midplane.

\begin{figure}[tpb]
\begin{center}
\includegraphics[clip=true,width=0.49\textwidth]{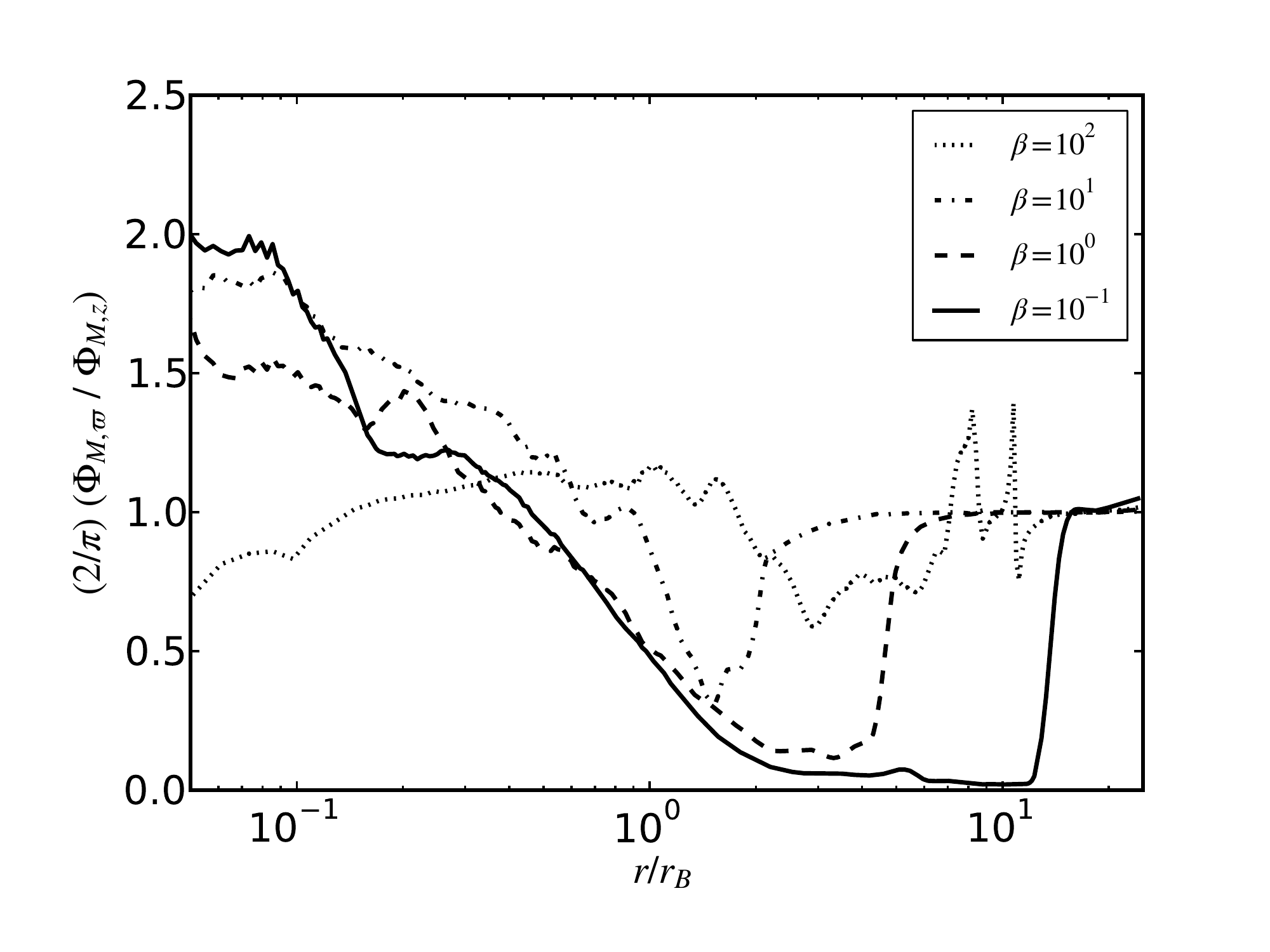}
\end{center}
\caption{The ratio of the mass influx in the equatorial direction to
the mass influx in the polar direction for several magnetized
models at $t = t_{\rm end}$, scaled so that uniform spherical inflow takes a
value of unity. \label{f2}}
\end{figure}

Accumulation of mass in the midplane is accompanied by a corresponding
increase in the inward gravitational attraction.  The magnetic flux
tubes that thread the disk are gradually pulled further toward the
accreting particle as the accumulation of mass in the midplane
continues.  We support this picture more quantitatively in Figure
\ref{f2}.  We use $\varpi$ to denote the cylindrical radius and plot
the ratio of the mass influx in the equatorial direction
\begin{equation}
\Phi_{M,\varpi} = \int_S \rho \vec{v} \cdot \hat{\varpi} \sin \theta d\theta d\phi
\end{equation}
to the mass influx in the polar direction
\begin{equation}
\Phi_{M,z} = \int_S \rho \vec{v} \cdot \hat{\vec{z}} \sin \theta d\theta d\phi
\end{equation}
along a spherical control surface $S$ of radius $r$ for each of the
magnetized models at $t = t_{\rm end}$.  The curves in Figure
\ref{f2} have been scaled by a constant $2/\pi$ so that uniform
spherical inflow takes a value of unity.  At large distances
($|\vec{x}| > \rb$), the flows become increasingly dominated by polar
inflow with increasing initial magnetic field strength.  However, at
smaller distances ($|\vec{x}| < \rb$), the cylindrical to polar influx
asymptotes toward $\sim 2$ with increasing magnetic field strength.
On smaller scales where magnetic forces break spherical symmetry, the
mass influx is predominantly along the equator.

As infall in the midplane proceeds, flux tubes that reach the
accreting particle are instantaneously liberated from the accreted
mass and accompanying gravitational force anchoring them.  This causes
episodic releases of strong, outward propagating flow.  This
configuration of outflow driven by magnetic buoyancy is known as the
magnetic interchange instability
\citep{Bernstein,furth}.  In the models with moderate
or strong initial magnetic fields strengths, corresponding to $\beta =
10$, $\beta = 1$ and $\beta = 0.1$, interchange unstable flows
originating at $r_{acc} \ll \rb$ lead to episodes of net outflow out
to radii comparable $\rb$ in the equatorial plane.  Flux tubes that
are outwardly released by resistive accretion are prevented from
escaping completely by the continued accretion pressure of the
surrounding gas.  The net mass inflow in these models is therefore
mediated by the rate at which inflowing material percolates through
this non-axisymmetric network of magnetically buoyant flow close to the
accreting particle.

\begin{figure}[tpb]
\begin{center}
\includegraphics[clip=true,width=0.49\textwidth]{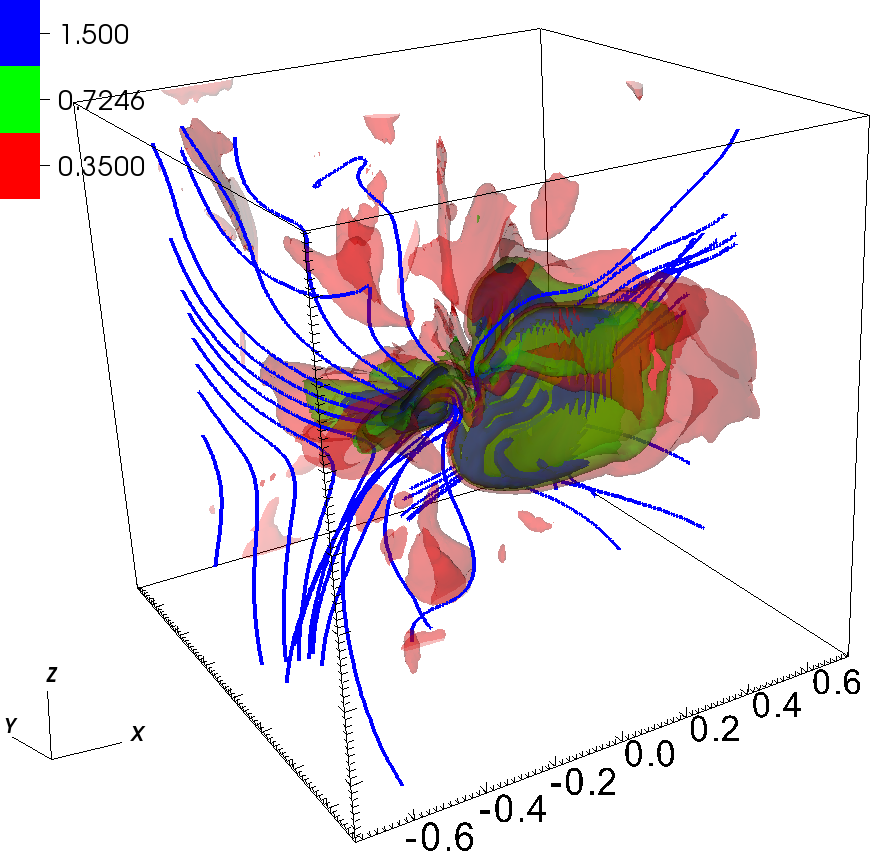}
\end{center}
\caption{Isosurfaces showing the innmermost $(1.5 \rb)^3$ of the
  magnetic shear parameter $\chi_{\rm mag}$ at $t = t_{\rm end}$ for the $\beta=100$ model
  indicating regions of magnetic reconnection due to tearing mode
  instability.  Blue curves represent magnetic field lines with
  footpoints evenly spaced along the y coordinate axis. \label{f3}}
\end{figure}

The models attain magnetic forces that balance $F_g$ at $r \sim \rb/2$
in the midplane by the time steady accretion sets in, independent of
the initial $\beta$.  The weak magnetic field lines in the $\beta =
100$ case become highly stretched before they are strong enough to
provide any resistance to being swept further inward as shown in the
top row of Figure \ref{f1}.  This flow leads to the development of
strong, thin current sheets and oppositely directed magnetic field
lines that closely approach each other in the midplane.  This
configuration is unstable to reconnection in magnetic resistive
tearing modes \citep{furth,rutherford}.  In the case of our numerical
code (and all ideal MHD codes), resistive reconnection occurs when
oppositely directed magnetic flux tubes become separated by $\lesssim
\Delta x$ and unresolved.  While the size scale of the ``magnetic
islands'' generated through this process is determined by the
numerical zone size, our numerical resolution is adequate to be sure
that this size scale is small compared to dynamical scale associated
with thermal ($\Delta x \ll r_{\rm B}$) and magnetic ($\Delta x \ll
\rab$) force gradients.  Furthermore, we have carried out resolution
studies to ensure that the resoultion used in our models is sufficient
to yield a converged late-time accretion rate.  Ultimately, mass
inflow is limited by the rate of production of magnetically isolated
islands by tearing mode reconnection in regions characterized by thin,
strong current sheets.  These islands continue toward the accreting
particle, unconnected to the global magnetic field structure.  As a
means to visualize flows that are most susceptible to reconnection by
numerical resistivity, we define the magnetic shear parameter
\begin{equation}
\chi_{\rm mag} = \frac{\Delta \vec{x} \cdot  (\grad \times \vec{B})}{|\vec{B}|}.
\end{equation}
Regions near or exceeding a magnetic shear parameter of $\sim 1$ are
highly susceptible to reconnection via magnetic tearing modes.  Figure
\ref{f3} gives a three dimensional sense of the geometry and scale of
the flows subject to numerical reconnection by plotting isosurfaces of the
magnetic shear parameter at $t=t_{\rm end}$, indicating efficient numerical
reconnection on scales of $r \lesssim \rb/2$.  Reconnection events
release magnetic tension that leads to magnetically tangled,
non-axisymmetric flow in this region.

\subsection{Comparison to Analytic Predictions for High $\beta$ Flow}
\begin{figure}[tpb]
\includegraphics[clip=true,width=0.49\textwidth]{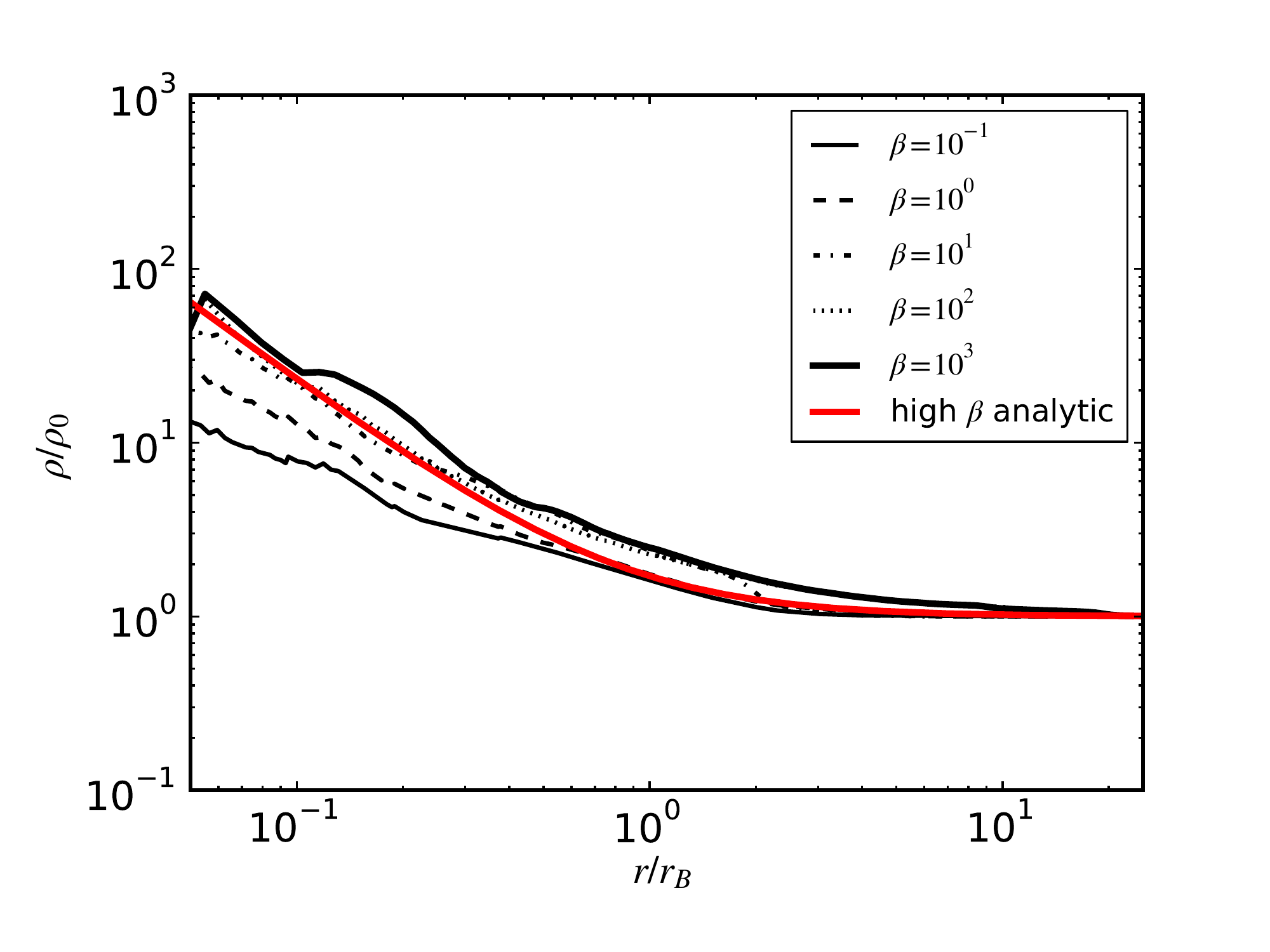}
\includegraphics[clip=true,width=0.49\textwidth]{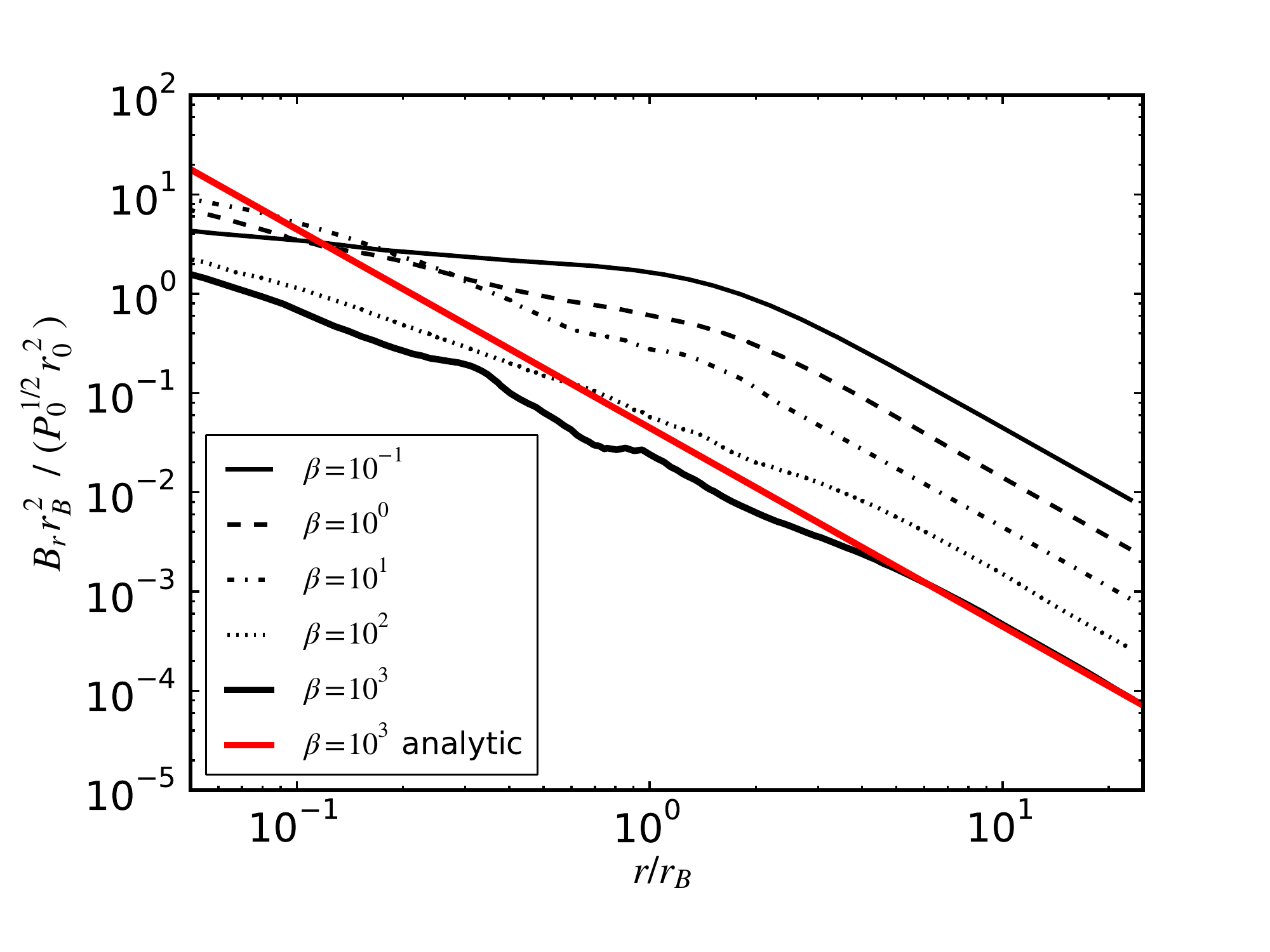} \\
\includegraphics[clip=true,width=0.49\textwidth]{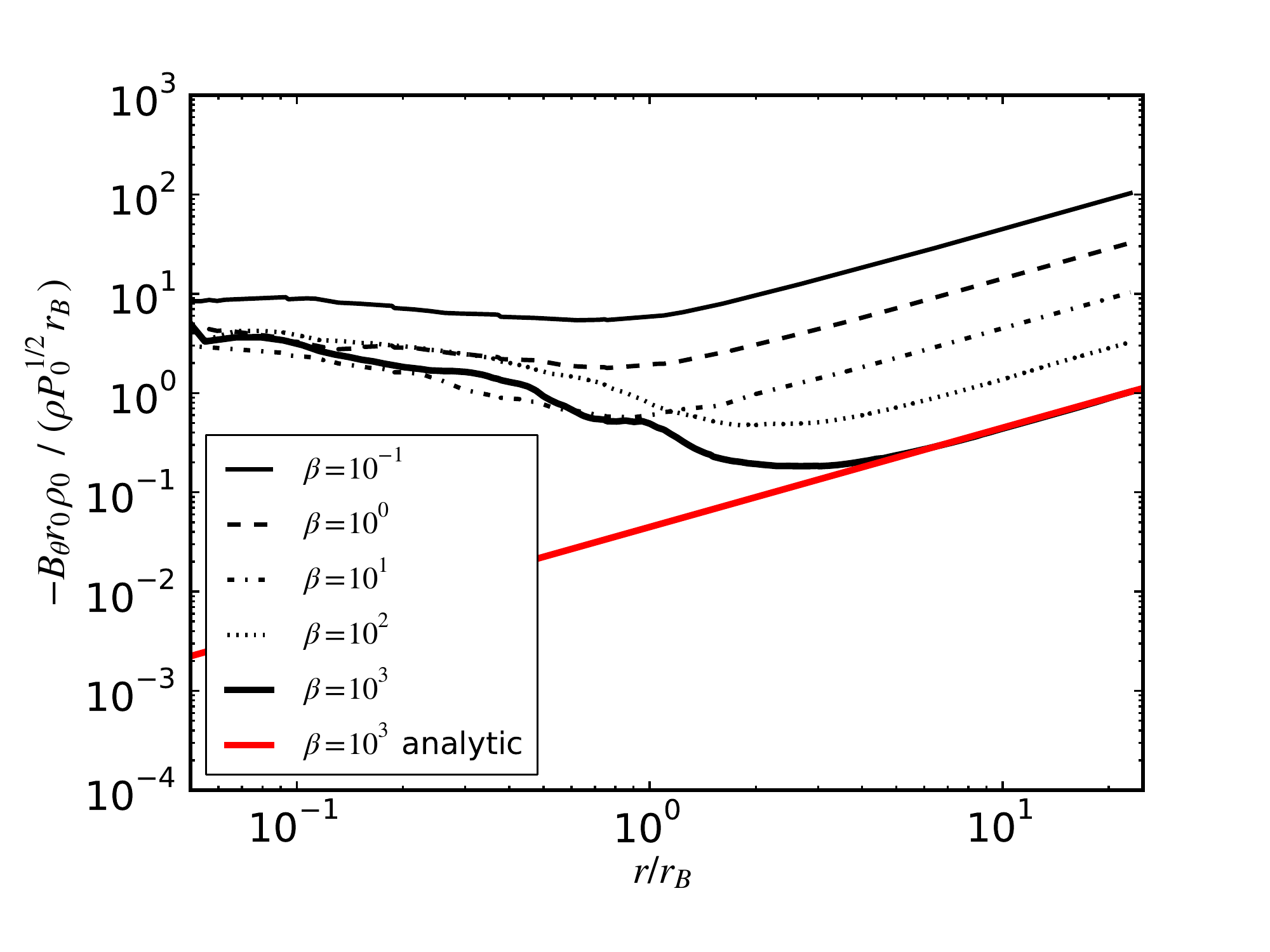}
\caption{Top Left: Azimuthally averaged density in the $z=0$ plane.
  Top Right: Azimuthally averaged radial component of the magnetic
  field in the $z=0$ plane, scaled inversely to the square of $r_0$.
  Bottom Left: Perpendicular component of the magnetic field along the
  $x=0,~ y=0$ axis, scaled by $r_0$ and inversely to the density.
  Each plot shows the analytic prediction for the limiting case of
  weak magnetic field with $\beta=1000$.  All of the plots are taken at
  time $t = t_{\rm end}$.
\label{f4}}
\end{figure}

Analytic predictions of the behavior of the accretion flows for the
limiting case of dynamically weak magnetic field are derived in
appendix \ref{A}.  The focus of this section is to compare the results
of the $\beta=100$ numerical model with these analytic predictions.
Equations (\ref{eq:rho}), (\ref{A23}) and (\ref{A25}) give predictions
of the steady state gas density, radial magnetic field and non-radial
magnetic field respectively.  (Results for the accretion rate will be
discussed in \S\ref{accretion}) It should be emphasized that $r_0$ in
these expressions is the initial position of gas that is at $r$ at
time $t$, and it must be evaluated numerically through the
transcendental equation (\ref{eq:r0}).  In Figure \ref{f4} we compare
these analytic predictions to the results of each of the magnetized
numerical models at $t = t_{\rm end}$.  The gas density, $\rho$, and
the non-radial magnetic field, $B_\theta$, are extracted from the
numerical models as azimuthal averages in the midplane of the
numerical domain where the sine term appearing in equation (\ref{A25})
is unity.  Likewise, the radial magnetic field $B_r$ is extracted from
the numerical models along the $x = y = 0$ axis where the cosine term
in equation (\ref{A23}) is unity.  The assumption of dynamically weak
magnetic field is met for $r \gtrsim \rb$ in the $\beta=1000$ model
and we find good agreement between the $\beta=1000$ model and the
analytic prediction at distances not too close to the origin.  The
analytic theory also agrees with the results for stronger fields for
$r \ga 4 \rb$.

In appendix \ref{A1}, equation (\ref{accflux}), we derive an analytic
prediction for the total magnetic flux that reaches the accretion
zone, $\Phi_a$, under the assumption of dynamically weak magnetic
fields and neglecting any possible reconnection that occurs near the
accretion zone. We have assumed that this flux escapes from the
accretion zone.  Even with reconnection, this method accurately tracks
the amount of escaping flux, although the time at which the flux
escapes may be altered by the reconnection.  Let $\Phi_\esc(r)$ be the
magnetic flux that is inside a radius $r$ and that has escaped from
the accretion zone. This quantity is well defined only for ideal MHD,
so that $r$ must be outside the region where magnetic reconnection
occurs.  At large values of $r$, $\Phi_\esc(r) \rightarrow \Phi_a$,
the total flux released during accretion.  As discussed above,
reconnection occurs in the inner regions of the flow, where it becomes
very turbulent. Outside this region, the flow is approximately
axisymmetric. There we can define $r_0$ as the initial radius of the
gas and magnetic flux, which at time $t$ is located in the midplane at
radius $r<r_0$. The initial flux inside $r_0$ is then the sum of the
flux inside $r(r_0)$ plus the flux that has escaped beyond $r$,
\begin{equation}
  \Phi_0[r_0(r)]=\Phi(r) + [\Phi_a-\Phi_\esc(r)],
  \label{eq:phiz}
\end{equation}
where $\Phi_0[r_0(r)] = |\vec{B_o}| \pi r_0^2$.  Equation (\ref{eq:r0})
gives $t$ as a function of $r$ and $r_0$; this can be inverted
numerically to obtain $r_0(r,t)$.  We note that equation (\ref{eq:phiz})
applies only outside the reconnection zone. If we had not assumed that
the flux could escape from the accretion zone after losing some of its
mass, flux would be conserved and both $\Phi_a$ and $\Phi_\esc$ would
vanish.

\begin{figure}[tpb]
\begin{center}
\includegraphics[clip=true,width=0.49\textwidth]{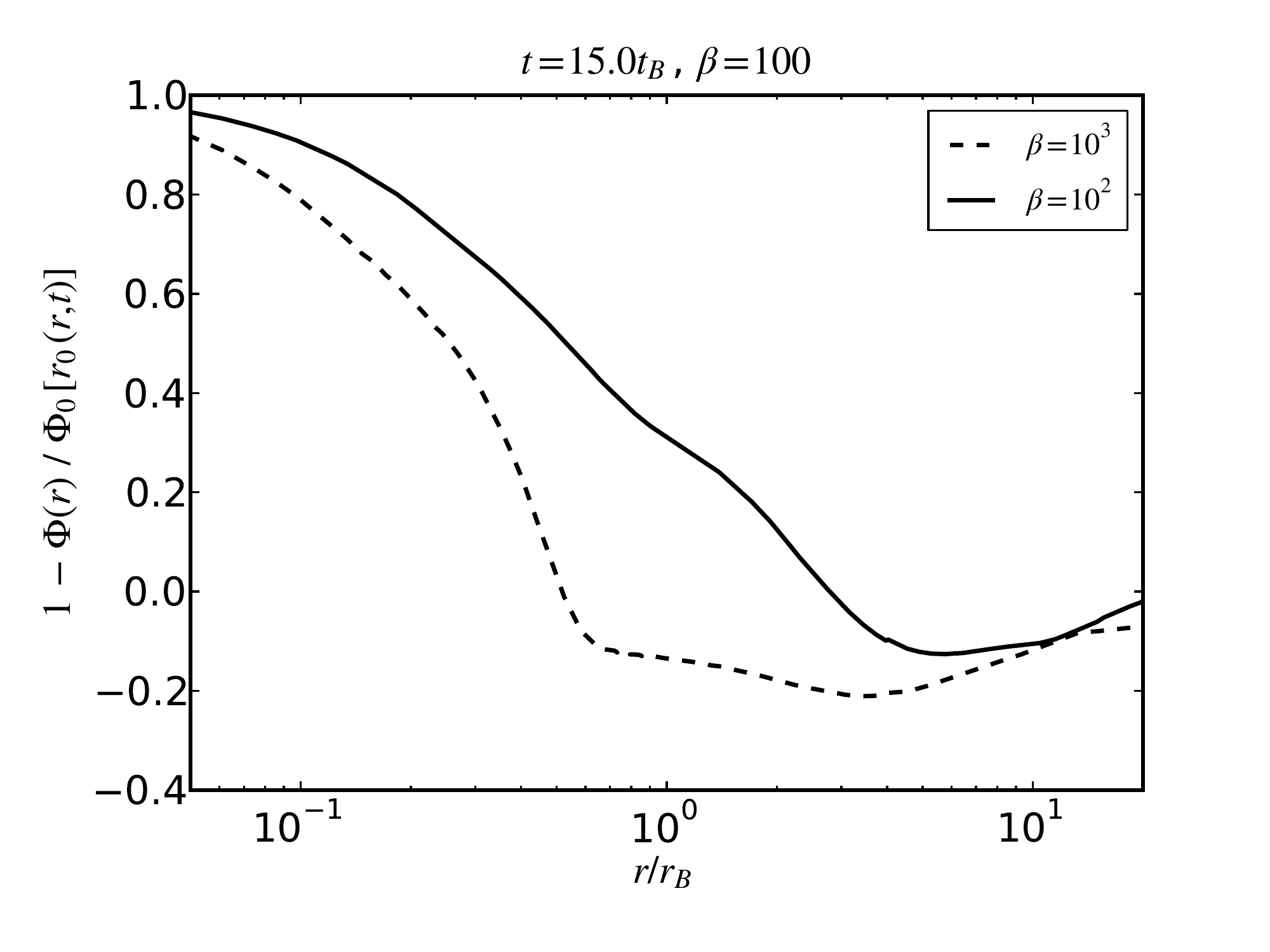}
\includegraphics[clip=true,width=0.49\textwidth]{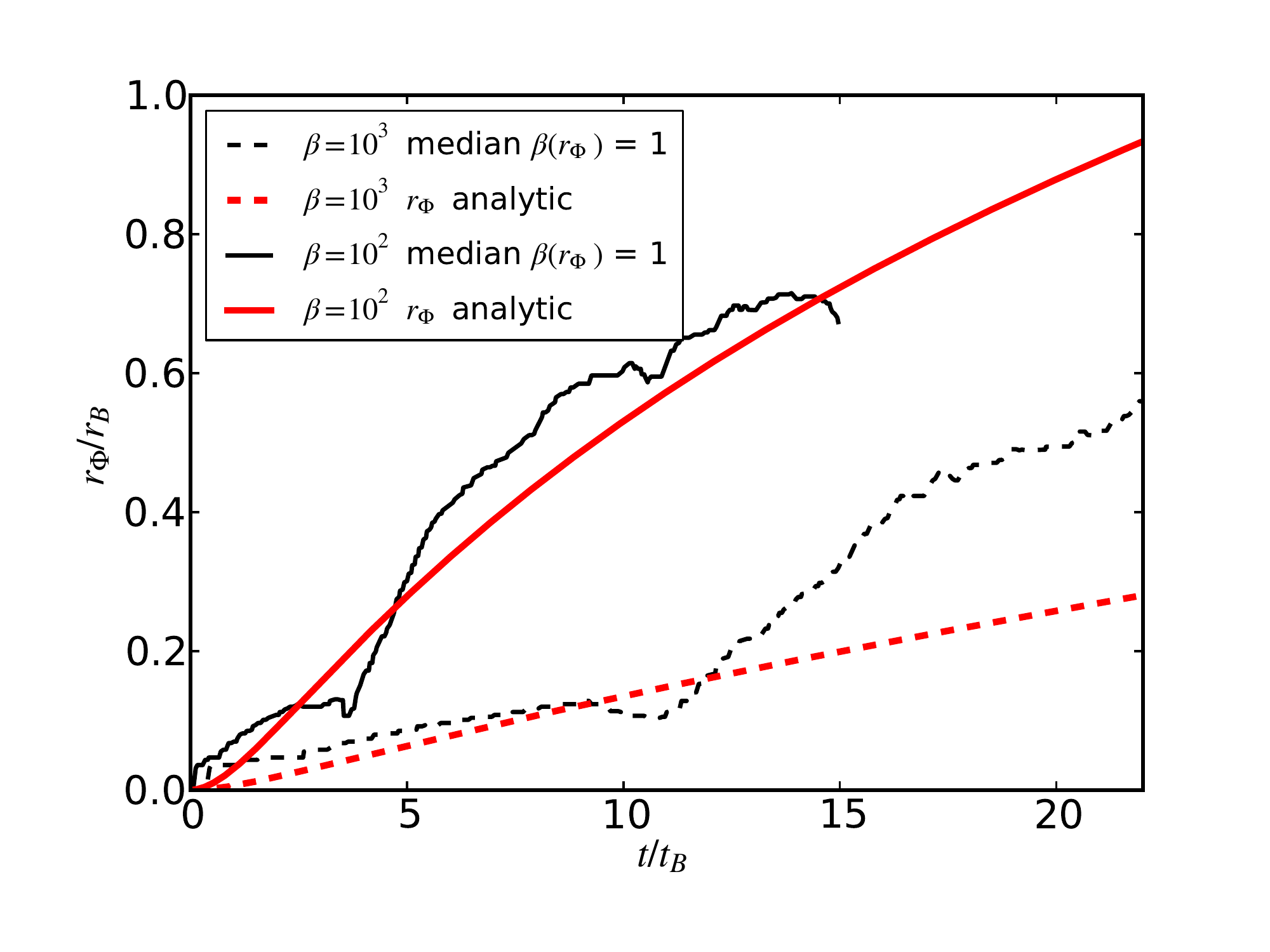}
\end{center}
\caption{Left: Radial distribution of escaped flux in the $\beta=100$
  and $\beta=1000$ models at time $t_{\rm end}$.  Right: The radial
  extent of the magnetically dominated region compared to the analytic
  prediction. \label{f5}}
\end{figure}

We can use our numerical models to test the predicted value of
$\Phi_a$ and to determine the radial distribution of the escaped flux.
To do this, we extract $\Phi(r)$ from our numerical result at a late
time ($t=15 \tb$), and we compare to the analytic result by rewriting
equation (\ref{eq:phiz}) as
\begin{equation}
  \delta_\Phi = \frac{\Phi_a-\Phi_\esc(r)}{\Phi_0[r_0(r)]} = 1-\frac{\Phi(r)}{\Phi_0[r_0(r)]},
\end{equation}

which is the fraction of the escaped flux that is beyond $r$.  In the
left panel of Figure \ref{f5} we show the above expression for the
high $\beta$ models.  In this case, the assumption of dynamically weak
magnetic fields used to derive the analytic estimate for $r(r_0)$ is
well met at $r \ga \rb$.  We expect that $\Phi(r)\approx
\Phi_0[r_0(r)]$ for $r\gg\rb$, and this is confirmed to within $10\%$
for $r > 4 \rb$.  Given our assumption of a resistive accreting
particle, we expect that $\Phi(r)\rightarrow 0$ as $r\rightarrow 0$,
and Figure \ref{f5} confirms this expectation by showing $\delta_\Phi
\rightarrow 1$ as $r \rightarrow 0$.  Furthermore, the accumulated
flux near the accreting particle shows strong evidence of escape for
$r \lesssim 1$, consistent with the scale of reconnection-driven
tearing modes shown in Figure \ref{f3} and discussed in section
\S\ref{morphology}.  The fact that $\delta_\Phi$ is greater than unity at
large radii is presumably due to the approximation made in determining
$r_0(r)$.  In the case of $\beta=100$, it appears that a significant
fraction of the escaped flux ($\ga20\%$) has moved outside $\rb$.

In appendix \ref{A1} we also predict the radius $r_\Phi$ out to which
the magnetic forces associated with the accumulated flux strongly
affect the flow. The analytic estimate of $r_\Phi$ for high $\beta$
flow is given by equation (\ref{eq:rphi}).  In the right panel of
Figure \ref{f5}, we plot this prediction against the radius where the
median plasma $\beta$ exceeds unity along the perimeter of a control
circle in the midplane of the high $\beta$ models.  At the latest time
shown, the prediction agrees with the simulation to within about 20\%
for the $\beta=100$ case. Note that at late times, the analytic
approximation has $r_\Phi\propto t^{1/3}$, but it is not known whether
the numerical results will continue to increase for $t>t_{\rm end}$.
It is not entirely clear why the $\beta=1000$ results do not agree
with the approximate model as well as the $\beta=100$ results. The
model predicts that $r_\Phi$ should be very close to (and slightly
less than) $r_{\Phi,1}$, given by equation (\ref{eq:rpo}) for
$\beta=1000$, whereas the simulations show that it is between
$r_{\Phi,1}$ and $r_{\Phi,2}$, given by equation
(\ref{eq:rpt}). This may be associated with the fact that the
escaped flux has gone well beyond the sonic point at $t_{\rm end}$ for
$\beta=1000$ (see the left-hand panel of Figure \ref{f5}), so that the
conditions are closer to those assumed in deriving $r_{\Phi,2}$ than
for $r_{\Phi,1}$.

\subsection{Accretion Rate}\label{accretion}
\begin{figure}[tpb]
\begin{center}
\includegraphics[clip=true,width=0.49\textwidth]{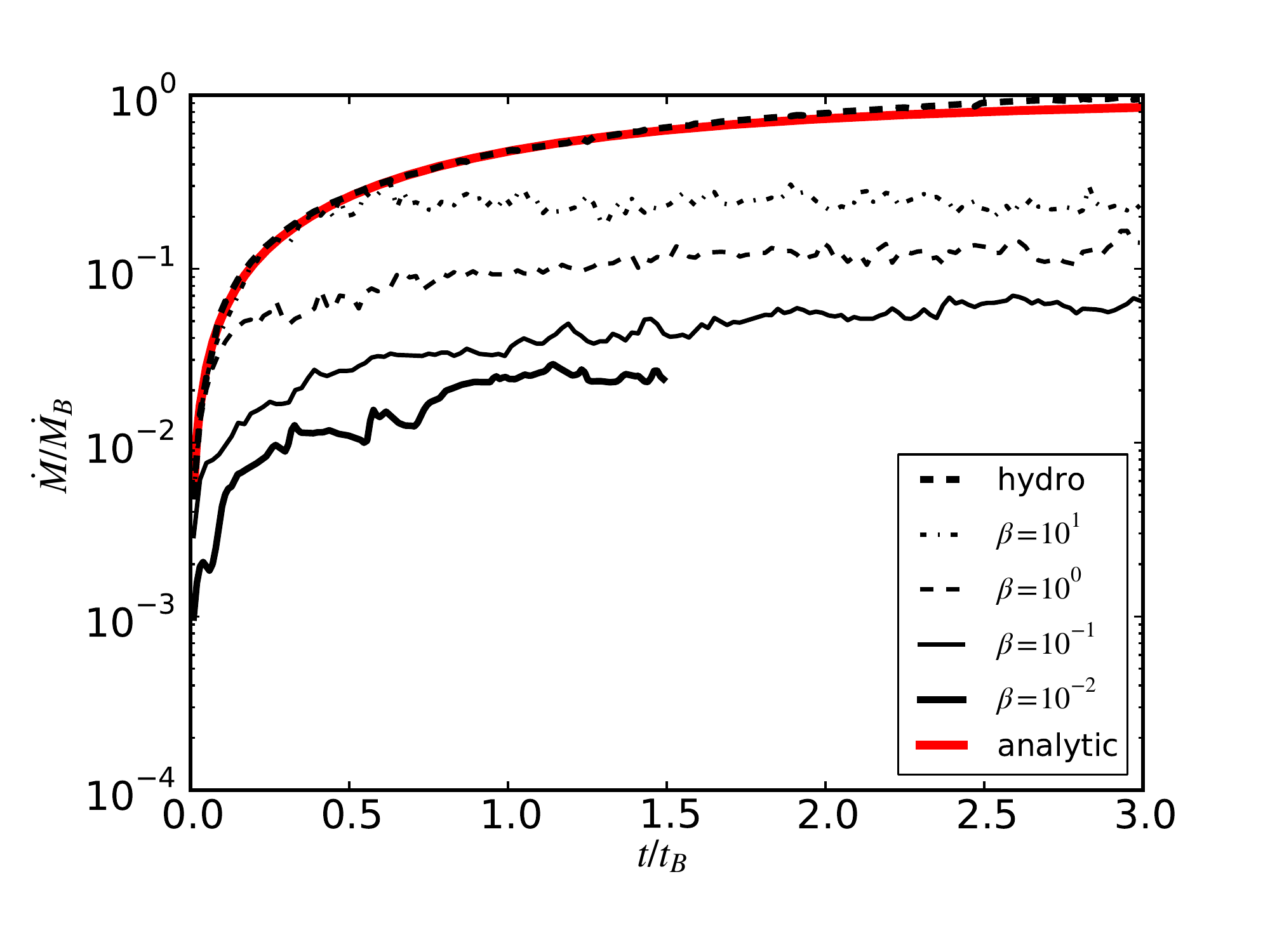}
\includegraphics[clip=true,width=0.49\textwidth]{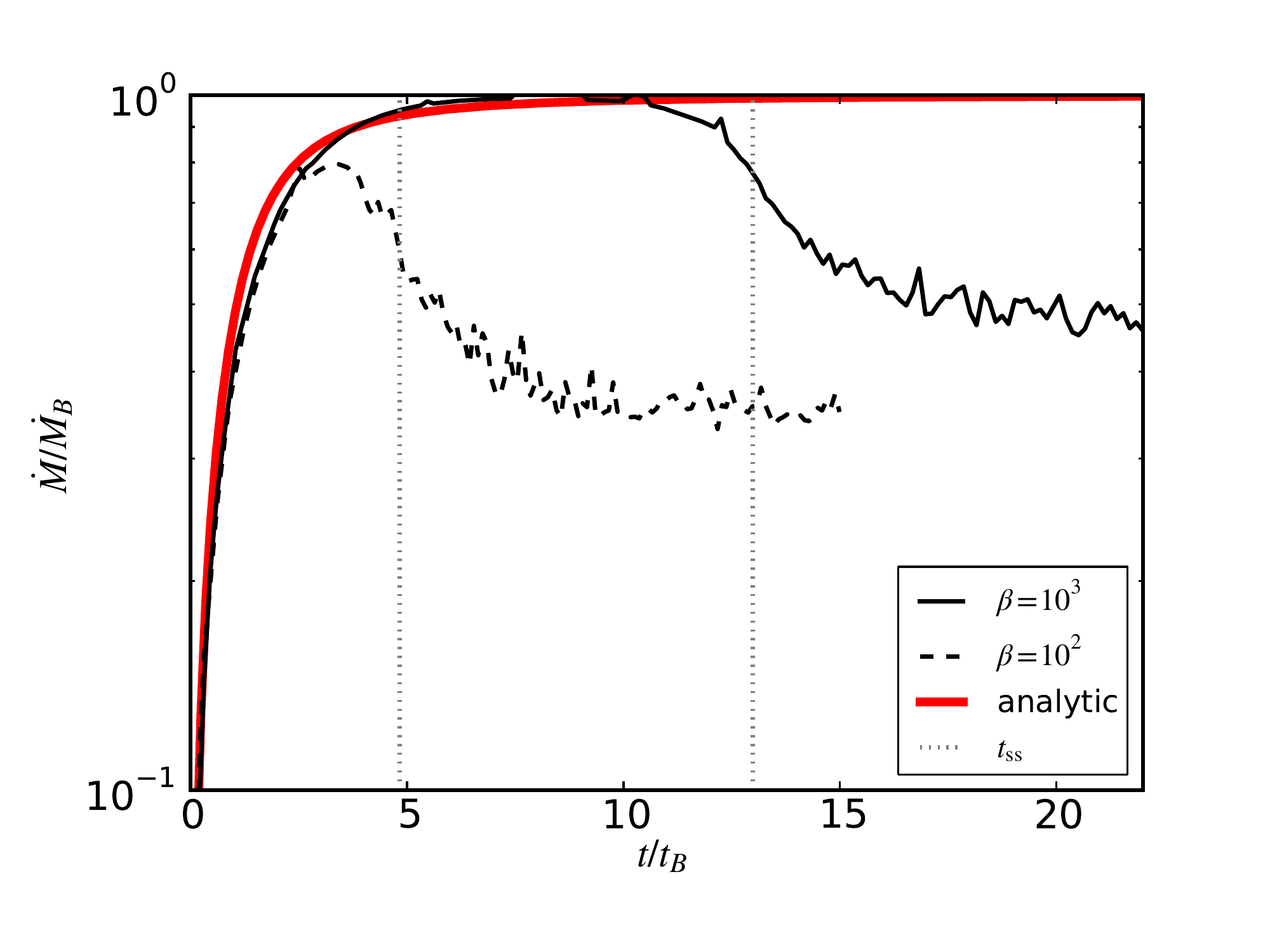}
\end{center}
\caption{Accretion rate as a function of time for each of the
  numerical models compared to the analytic prediction for the
  time-dependent accretion rate for the purely hydrodynamic case. In
  the right plot the time $t_{\rm ss}$ indicated by a gray vertical
  line when the accretion rate is midway between the maximum accretion
  rate and the final steady state accretion rate, representing the
  characteristic time for the flow to transition from Bondi accretion
  to a magnetically mediated steady state.  \label{f6}}
\end{figure}

Figure \ref{f6} shows the rate of accretion onto the central particle
as a function of time for each of the numerical models.  The left plot
also includes the result of a purely hydrodynamic control model for
comparison.  As discussed in \S\ref{morphology}, the magnetized models
reach a statistically steady accretion rate with inflow mediated by
reconnection and/or the interchange instability, whereas
the purely hydrodynamic model asymptotically approaches the truly
steady, spherical Bondi flow.  The high frequency modes in Figure
\ref{f6} have been smoothed using a box-car smoothing width of $0.02
\tb$.  The red dashed curve shows the analytic approximation for the
time-dependent accretion rate without magnetic fields from appendix
\ref{A1}, equation (\ref{eq:mdot}).  The analytic estimate is in
excellent agreement with the purely hydrodynamic numerical model.

\begin{figure}[tpb]
\begin{center}
\includegraphics[clip=true,width=0.5\textwidth]{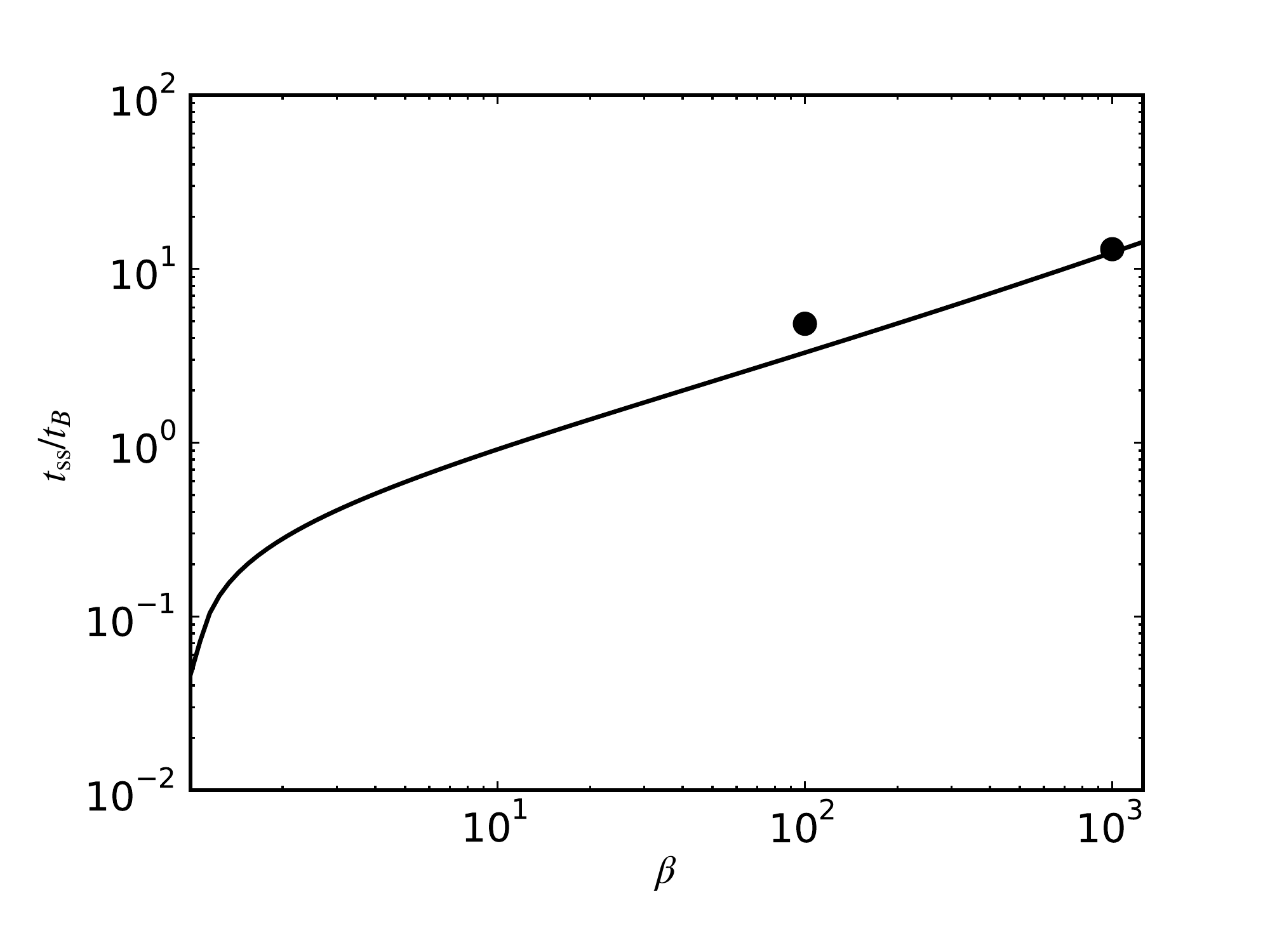}
\end{center}
\caption{An analytic estimate of the time required for enough magnetic
flux to accumulate inside of the thermal sonic radius for the flow to
reach a state of magnetically mediated accretion. Black circles
indicating the time when the $\beta=10^{2}$ and $\beta=10^{3}$
simulations transition from Bondi to magnetically mediated flow are in
good agreement with the analytic prediction.  \label{f7}}
\end{figure}

An interesting aspect of the results shown in Figure \ref{f6} is that
the weak magnetic field models ($\beta=100$ \& $\beta=1000$) undergo
an initial transient of rapid accretion before settling into a steady
accretion rate.  The reason for this is that enough time must elapse
for sufficient magnetic flux to accumulate close to the accreting
particle for the accretion to the surface of the particle to become
magnetically dominated, whereas thermal pressure dominates close to
the particle during the initial development of the flow.  We can use
equation (\ref{eq:r0}) to estimate the time required for the flow to
settle into a magnetically mediated steady state accretion regime.
Specifically, we estimate the time to reach this steady state, $t_{\rm
ss}$, as the time required for enough magnetic flux to accumulate
inside the thermal sonic radius, $r_{\rm sonic} = \rb/2$
\citep{bondi}, so that the average magnetic field within $r<r_{\rm
sonic}$ in the midplane corresponds to $\beta=1$ (i.e., $\bar
B=(8\pi\rho_0 c^2)^{1/2}$ for $r<r_{\rm sonic}$ at $t=t_{\rm ss}$).
Neglecting any flux that has escaped beyond $r_{\rm sonic}$, this then
implies
\begin{equation}
  \pi r_0(r=r_{\rm sonic},t=t_{\rm ss})^2  = \pi r_{\rm sonic}^2 \beta^{1/2}\label{t_ss}
\end{equation}
Solving this for $t_{\rm ss}$ using the transcendental expression for
$r_0$ in equation (\ref{eq:r0}) determines $t_{\rm ss}(\beta)$, as shown
in Figure \ref{f7}.  The simulations match with this prediction with
the $\beta=100$ and $\beta=1000$ models transitioning toward the
magnetically dominated steady state accretion rate at $t \sim t_{\rm
ss}$ as shown in Figure \ref{f6}.

\begin{figure}[tpb]
\begin{center}
\includegraphics[clip=true,width=0.5\textwidth]{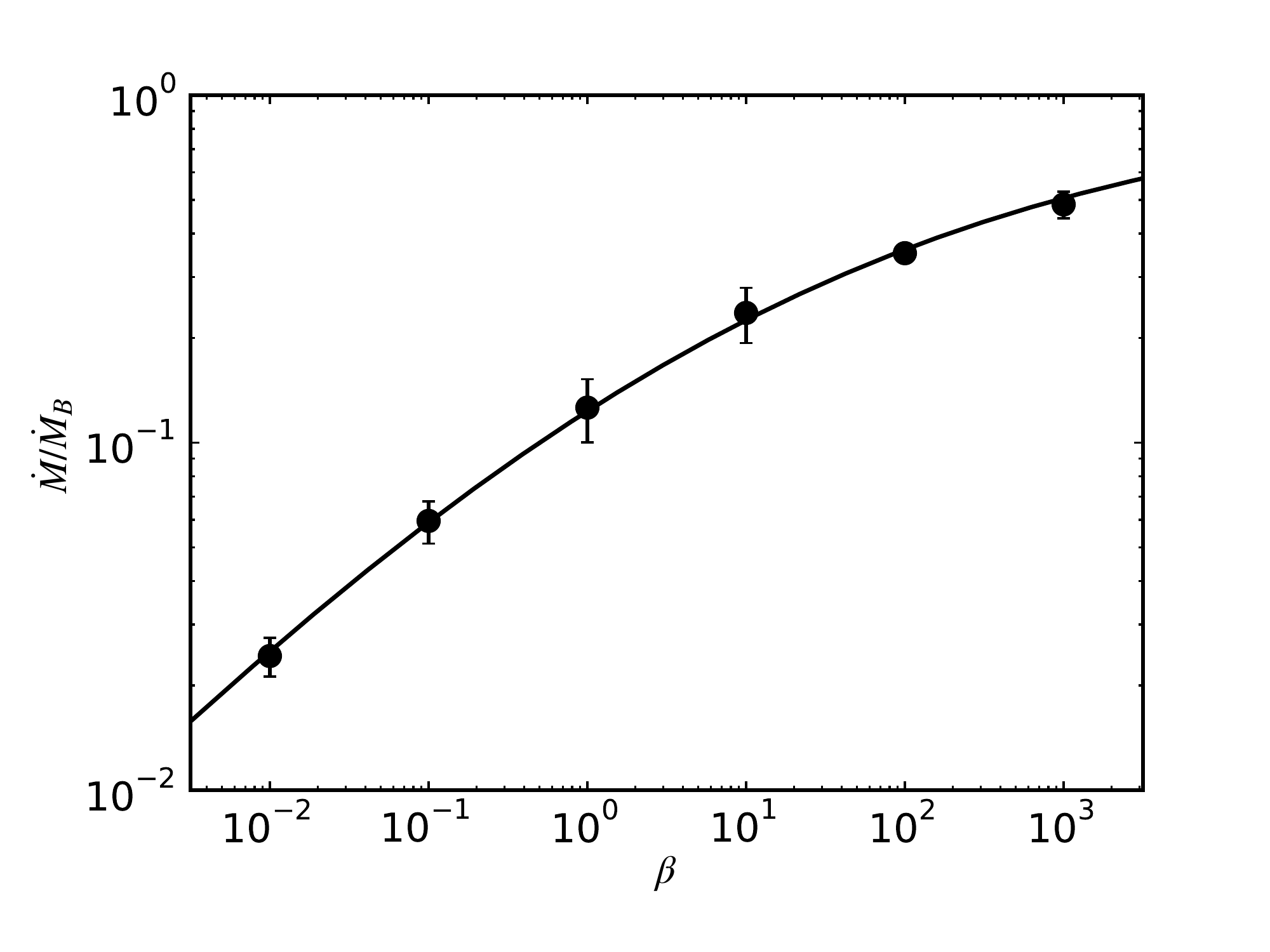}
\end{center}
\caption{Average accretion rate as a function of plasma $\beta$
  parameter.  Error bars show the standard deviation in the accretion
  rate due to interchange and tearing mode unstable flows near the
  accreting particle. The solid line shows equation (\ref{2param})
  with the best-fit coefficients $\beta_{\rm ch}=5.0$ and
  $n=0.42$. \label{f8}}
\end{figure}

Figure \ref{f8} shows the average accretion rate over the last $t_{\rm
  B}$ of the simulated time for each of the $\beta=10^{-1}$ -
$\beta=10^{3}$ models as black circles.  The $\beta=10^{-2}$ model was
run only to $t_{\rm end}=1.5 t_{\rm B}$ and for that case we average
over the last $t_{\rm B}/2$ of the simulated time.  The vertical bars
on each point indicate the standard deviation of the accretion rate
over the same time interval.  It should be noted that these should be
interpreted as a measure of the effect of small scale departure from
steady accretion flow due to MHD flow instability and not as ``error
bars'' in the usual sense of measurement uncertainty.  The accretion
rate data are presented in tabular form as well in Table \ref{t2}.

\begin{deluxetable}{l c c}
  \tablewidth{0pt} \tablecaption{Accretion Rates.\label{t2}}
  \startdata \tableline $\beta$ & $\dot{M} / \dot{M_{\rm B}}$ &
  $\sigma_{\dot{M}} / \dot{M_{\rm B}}$ \\ \tableline 1000 & 0.48 & 0.043 \\ 100 & 0.35 &
  0.015 \\ 10 & 0.24 & 0.043 \\ 1 & 0.13 & 0.26 \\ 0.1 & 0.060 & 0.083
  \\ 0.01 & 0.024 & 0.031 \\ \enddata \tablewidth{\textwidth}
  \tablecomments{ Second column: Normalized mean accretion rate for
    the isothermal equation of state models. Third column: Standard
    deviation of the isothermal accretion rate.}
\end{deluxetable}

We can obtain a simple analytic model for the accretion flow in the
magnetically dominated case by assuming that the gas flows in from the
\alfven\ radius $\rab$ at the \alfven\ velocity after collapsing
vertically from a distance of order the Bondi radius, $\rb$:
\begin{equation}
\dot M\propto 2\pi \rab\cdot 2\rab\cdot \rho_\infty \va
\propto \dot M_{\rm B} (c/\va)\propto\dot M_{\rm B} \beta^{1/2}~~~~(\beta\ll 1),
\end{equation}
where the second expression follows from equation (\ref{eq:ra}).  We
note that \cite{toropin} have shown similar accretion rate dependence
with magnetic pressure close to the accreteor for the case of the
accretion onto a magnetized star. We estimate of the constant of
proportionality in the above expression that is in rough agreement
with our numerical results by a more through analytic consideration of
the problem in Appendix \ref{B2}.  Since $\dot M\rightarrow \dot
M_{\rm B}$ at large $\beta$, a simple relation that captures the
limiting behavior in both cases is
\begin{equation}
\frac{\dot{M}}{\dot{M}_{\rm B}} = \left(\left[\frac{\beta_{\rm ch}}{\beta}\right]^{n/2} + 1\right)^{-1/n}. \label{2param}
\end{equation}
The solid line in Figure \ref{f8} is based on a least-squares fit in
$\log\beta - \log\dot M$ space for the parameters $\beta_{\rm ch}=5.0$
and $n=0.42$.  In this notation, $\beta_{\rm ch}=5.0$ gives the
characteristic value of $\beta$ for the transition from the high and
low beta limiting cases to occur.  

Sub-grid particle accretion methods have been employed to model
protostellar accretion in numerical simulations of protostellar cores
and clouds by several authors
\citep{bate,KrumholzSinks,federrath,wang,padoan}.  Equation
(\ref{2param}) should be of particular utility for extending the
sub-grid accretion model for embedding Lagrangian sink particles on an
Eulerian mesh of \cite{KrumholzSinks,KrumholzStars} and \cite{Offner}
to the magnetic case for particles moving subsonically through the
ambient medium.

It is noteworthy that the qualitative behavior we find at late times is remarkably similar to that discovered by \cite{krumholzvorticity} for the case of hydrodynamic Bondi accretion of a gas with vorticity. The Kelvin circulation theorem for a non-viscous flow is analogous to flux-freezing in ideal MHD \citep{ShuBook}, and in the problem of accretion from a vortical fluid, the dimensionless vorticity parameter $\omega_* \equiv |\nabla \times \mathbf{v}|/(c/r_B)$ defined by \cite{krumholzvorticity} is analogous to $\beta^{-1/2}$ in the present work.\footnote{The $-1/2$ power arises because the magnetic flux at infinity varies as $\beta^{-1/2}$, while the vorticity at infinity scales as $\omega_*$.} In both cases, the accretion flow causes a buildup of vorticity / flux near the accreting object, which produces regions where the outward centrifugal / magnetic force is able to balance gravity and inhibit accretion. For Bondi accretion with vorticity, flows with strong vorticity ($\omega_*\gg 1$) have steady-state accretion rates that scale as roughly $\omega_*/\ln \omega_*$, nearly identical to the $\beta^{-1/2}$ scaling we find for the strongly magnetized case ($\beta \ll 1$). For the weak vorticity case ($\omega_* \ll 1$), the accretion rate initially rises to nearly $\dot{M}_B$, but then declines as vorticity builds up, reaching an asymptotic value $<\dot{M}_B$ after a transient whose duration is proportional to $\omega_*^{-1}$. The high $\beta$ cases here behave in precisely the same way.

The only difference we can identify is that, in the vortical case, \cite{krumholzvorticity} find the accretion rate converges to a value slightly less than $\dot{M}_B$ in finite time, even in the limit $\omega_* \rightarrow 0$, as long as it is not so small as to place the circularization radius within the physical size of the accretor. Here we find that the accretion rate at time $t > t_{\rm ss}$ appears to converge to $\dot{M}_B$ as $\beta\rightarrow \infty$.\footnote{However, it is not clear from our simulations if the accretion rate would converge to $\dot{M}_B$ or some lower accretion rate in the limit of large $\beta$ at $t=\infty$ since then a finite flux could in principle build up near the particle.} The origin of the difference is not entirely clear, but one possibility has to do with mechanisms for removing excess vorticity / flux. Both can be removed by advection, but magnetic flux can also be rearranged by reconnection, as occurs in our simulations. In addition, magnetic buoyancy tends to cause regions of high flux to rise away from the accretor. (Similar effects are seen in simulations by \cite{vazquez11}.) In a non-viscous flow, there are no analogous processes capable of rearranging the vorticity. In real astrophysical systems, non-ideal MHD and magnetic bouyancy effects almost always occur at larger scales than those on which molecular viscosity becomes important, and this may lead to a real difference in behavior at late times in the weak vorticity / field cases.

\section{Comparison to Adiabatic Models}
It is illustrative to compare our accretion rates to those of earlier
works that considered the accretion of magnetized gases with a similar
field topology but an ideal gas law equation of state ($\gamma = 5/3$)
appropriate for accretion without radiative losses.  \cite{pang} found
that for $1<\beta<100$, the accretion rate in the adiabatic case
depends explicitly on the size of the accreting particle with
vanishing accretion rate as the particle size $\rightarrow 0$.  In
contrast, for the isothermal case, we find asymptotic convergence
toward a finite accretion rate with decreasing grid spacing and
particle size, even for cases with very strong large scale fields (see
Appendix \ref{convergence}).  In the case of adiabatic flow, the
results of \cite{pang} and \cite{igu} show that mass accumulation in
the midplane is limited by thermal pressure.  In addition, magnetic
reconnection leads to thermal pressure-driven convective flows that
also inhibit mass accumulation in the adiabatic case.  The work of
\cite{pang} has shown that at sufficiently small scale, these effects
completely halt accretion.  Because both of these effects are driven
by thermal pressure, neither of them appear in our simulations for the
isothermal regime.  Consequently, radiatively efficient Bondi-type
flows threaded by large-scale magnetic fields converge to a finite
accretion rate in the limit of vanishing accreting particle size.

\section{Conclusions}
We have carried out a numerical study of the effect of large-scale
magnetic fields in an isothermal gas on the rate of accretion onto a
resistive point mass---i.e., for the case in which only mass, not
magnetic flux, accretes onto the point mass. The assumption of
isothermality is approximately satisfied in regions of star formation,
where the cooling time of the molecular gas is generally much shorter
than the dynamical time for accretion.  The simulations for this study
use simple, very general initial conditions that avoid complications
arising from boundary conditions by keeping the boundaries far from
the accreting object.  At the same time, our simulations leverage the
AMR methodology to retain high spatial fidelity close to the accreting
object.  Contrary to the adiabatic case \citep{pang}, our simulations
show convergence toward a finite accretion rate as the radiius of the
accreting object vanishes, regardless of magnetic field strength.  We
find that magnetic fields reduce the Bondi accretion rate in an
isothermal medium by about a factor 2 for weak magnetic fields
(plasma-$\beta$ parameter $\ga 100$) at late times, when the magnetic
field near the point mass builds up to the point that it can impede
accretion. For strong fields ($\beta\ll1$), the accretion rate is
reduced by a factor $\sim \beta^{1/2}/4$.  We have developed
approximate fitting formulae for the accretion rate as a function of
$\beta$. 
The Appendixes give analytic results for the time dependent
accretion rate of a point mass in the limit of negligible magnetic
field and for the steady-state accretion rate for the case of a strong
magnetic field; both are in good agreement with the results of the
simulations.

\acknowledgements The authors are grateful of helpful discussions with
Eric Agol and Aaron Lee on the topic of this paper.  Support for this
work was provided by: the US Department of Energy at the Lawrence
Livermore National Laboratory under contract DE-AC52-07NA27344 (AJC
and RIK); an Alfred P.\ Sloan Fellowship (MRK); NASA through ATFP
grant NNX09AK31G (RIK, CFM, and MRK); the National Science Foundation
through grants AST-0807739 (MRK) and AST-0908553 (RIK and CFM); NSF
grant CAREER-0955300 (MRK) and NASA through a Spitzer Space Telescope
Theoretical Research Program grant (CFM and MRK). Support for computer
simulations was provided by an LRAC grant from the National Science
Foundation through TeraGrid resources and the NASA Advanced
Supercomputing Division.  LLNL-JRNL-497719

\appendix
\section{Numerical Convergence} \label{convergence}
\begin{figure}[tpb]
\includegraphics[clip=true,width=0.49\textwidth]{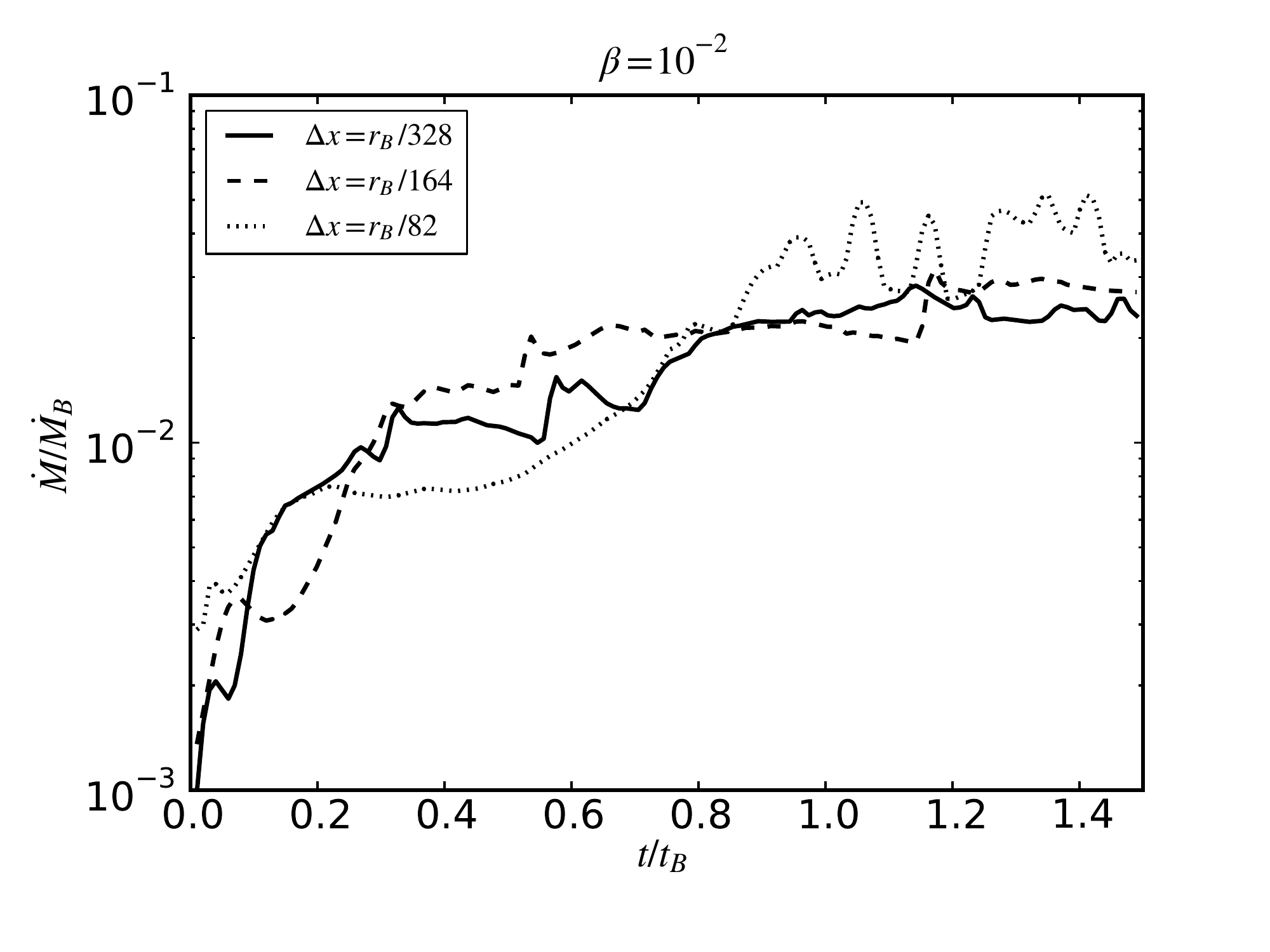}
\includegraphics[clip=true,width=0.49\textwidth]{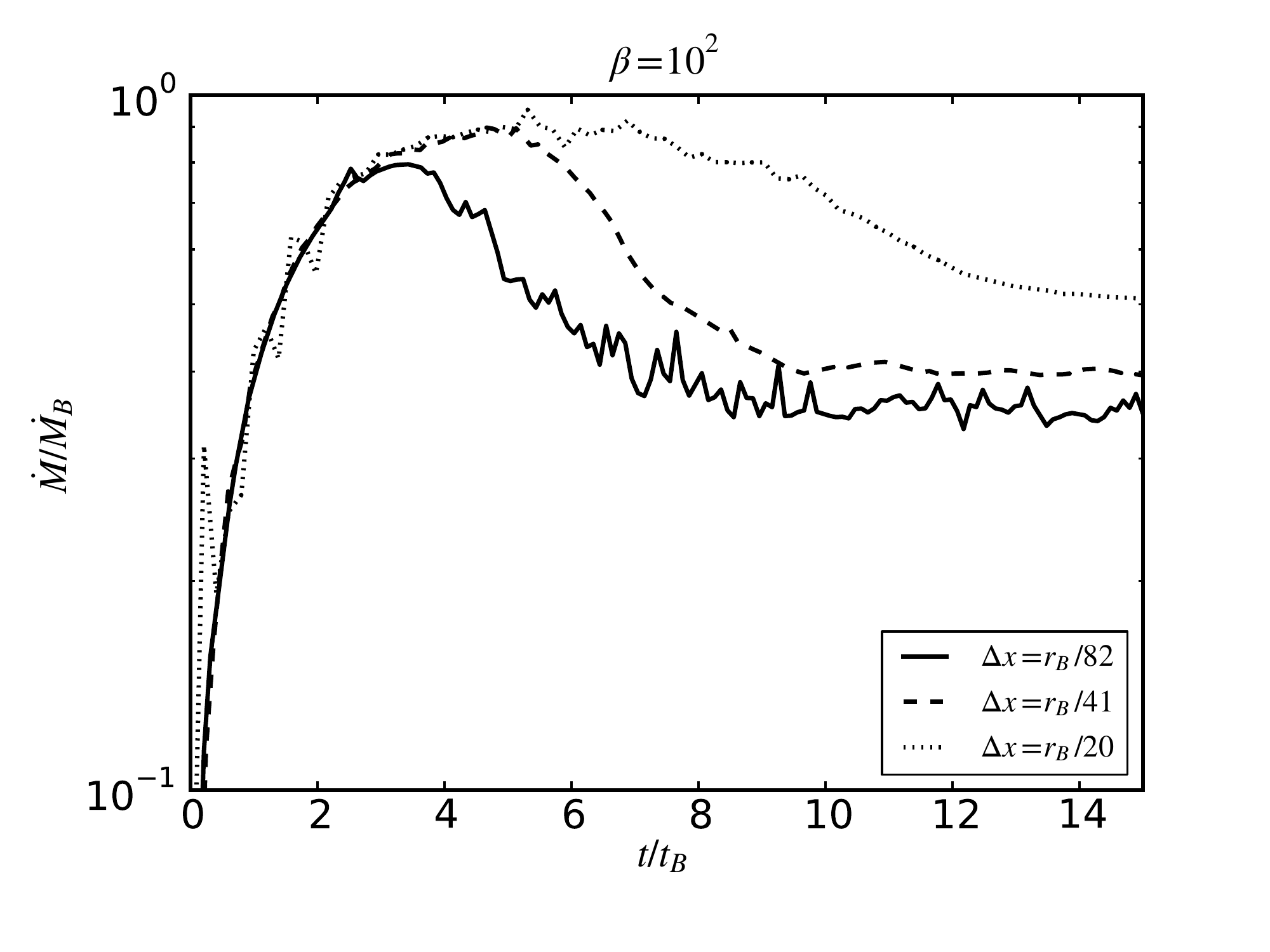}
\caption{Convergence properties of the accretion rate of selected
  models as a function of time.}  \label{f9}
\end{figure}

The mean steady-state accretion rate at late time is the principle
quantity of interest from the numerical models presented in this
paper.  In this section we demonstrate that our models provide
well-converged estimates for this result.  As discussed in
\S\ref{setup}, the \alfven-Bondi radius $\rab$ in our models becomes
less resolved as $\beta$ decreases, for a fixed numerical resolution
scale.  Additionally, our numerical models at $\beta=10^3$ and
$\beta=10^2$ use a coarser resolution than the lower beta models owing
to the computational constraints imposed by the longer simulation time
required to achieve steady accretion.  We therefore focus on
demonstrating convergence for the set of models that are least
resolved in $\rab$, namely, the model with the strongest magnetic
field ($\beta=10^{-2}$) at resolution $\Delta x =
328~\textrm{zones}/\rb$ and the model with the strongest magnetic
field ($\beta=10^2$) at the resolution of $\Delta x =
82~\textrm{zones}/\rb$.  Figure \ref{f9} shows the time-dependent
accretion rate for each of these models at their native resolution,
with the resolution and effective sink particle radius coarsened by a
factor of 2 and with the resolution and effective sink particle radius
coarsened by a factor of 4.  The convergence properties of the
instantaneous accretion rate at any particular time is difficult to
assess owing to the stochastic nature of the accretion rate.  However,
we can assess the convergence properties of the accretion rate
averaged over a time interval that is sufficiently long to diminish
the impact of these stochastic effects.  We choose $\tb/2$ for the
$\beta=10^{-2}$ model and $t_b$ for the $\beta=10^2$ model.

We find that the late time averaged accretion rates at the resolutions
shown in Figure \ref{f9} exhibit asymptotic convergence with an implied
order of accuracy
\begin{equation}
p = \frac{1}{\ln(2)} \ln\left(\frac{\dot{M}_{2x}-\dot{M}_{4x}}{\dot{M}-\dot{M}_{2x}}\right) \label{order}
\end{equation}
that is better than first order accurate.  In equation (\ref{order}),
$\dot{M}$ is the time-averaged accretion rate at the native
resolution, $\dot{M}_{2x}$ is the time-averaged accretion rate at a
resolution that is coarsened by a factor of 2 and $\dot{M}_{4x}$ is
the time-averaged accretion rate at a resolution that is coarsened by
a factor of 4.  Given the shock-capturing nature of the {\ttfamily RAMSES} code,
we cannot guarantee that better than first order convergence would
continue at even higher resolution.  We therefore estimate the
numerical grid convergence error using Richardson-extrapolation under
the conservative assumption of a first order rate of convergence as
\begin{equation}
\epsilon = \left|\frac{\dot{M}-\dot{M}_{2x}}{\dot{M}}\right|
\end{equation}
In Table \ref{t3} summarizes the convergence properties for each of
the models considered in this section.  We find that the time-averaged
accretion rates given by our native resolution numerical models are
accurate to within $14\%$ of the Richardson-extrapolation estimate of
the asymptotically converged result.

\begin{deluxetable}{l l l l l l}
  \tablewidth{0pt}
  \tablecaption{Convergence Properties.\label{t3}}
  \startdata
  \tableline
  $\beta$ & $\dot{M}_{4x}$ & $\dot{M}_{2x}$ & $\dot{M}$ & $p$ & $\epsilon$ \\
  \tableline
  100  & 0.0513 & 0.0400 & 0.0351 & 1.20 & 0.14 \\
  0.01  & 0.0380 & 0.0258 & 0.0243 & 2.46 & 0.062 \\
  \enddata
  \tablewidth{\textwidth}
\end{deluxetable}

\section{Bondi Flow with a Weak Magnetic Field}\label{A}

\subsection{Dynamics}\label{A1}

Here we calculate Bondi flow under the assumption that the gas density is
initially uniform and then evolves into a steady state. This initial condition
corresponds to that in our numerical simulations, but would be difficult to
realize in practice (for example, an approximation
to this situation might result when gas flowing supersonically
past an object is suddenly brought to rest by a strong shock).
We assume that the magnetic field is weak so that it does not affect
the flow. As we shall see below, this
approximation breaks down sufficiently close to the central mass or at sufficiently
late times. The flow is then spherically symmetric, and in a steady state the accretion rate is
\beq
\dot M=4\pi\lambda\rb^2\rho_\infty c ,
\label{eq:dotms}
\eeq
where
\beq
\rb\equiv \frac{GM_*}{c^2}
\eeq
is the Bondi radius associated with a star of mass $M_*$ and
 $\lambda \simeq 1.1$ for isothermal flow.
For a steady accretion flow, we then have
\beq
4\pi r^2 \rho v=4\pi \rb^2\rho_\infty c.
\label{eq:steady}
\eeq 
At large radii ($r\gg \rb$), we have $\rho\simeq\rho_\infty$ so that
\beq
\frac{v}{c}\simeq \frac{\rb^2}{r^2}.
\label{eq:vc}
\eeq
Henceforth, we shall normalize lengths to $\rb$, velocities to $c$, and
times to $\rb/c$; equation (\ref{eq:vc}) then becomes $v=r^{-2}$.
If we assume that the mass element is initially at rest at $r_0$, then
at small radii or at early times, the gas is in free fall, so that
\beq
v=\surd 2\left(\frac{1}{r}-\frac{1}{r_0}\right)^{1/2}.
\label{eq:vc2}
\eeq
An approximation for the flow everywhere is
\beq
\frac{1}{v}\simeq \frac{1}{\surd 2}\left(\frac{1}{r}-\frac{1}{r_0}\right)^{-1/2} + r^2.
\label{eq:ov}
\eeq
It should be noted that, although we used the approximation of steady flow to
estimate the velocity at large radii, equation (\ref{eq:ov}) for the velocity is time dependent:
$r_0$ is a function of both $r$ and $t$, so $\partial v/\partial t \neq 0$. In equation
(\ref{eq:mdot}) below we shall give the time-dependent result for Bondi flow that occurs
in an initially stationary medium.

How long does it take a particle to reach a point $r$ when it starts at $r_0$? Integration
of equation (\ref{eq:ov}) gives
\beqa
t&=&\int_r^{r_0} \frac{dr}{v},\\
&=& \frac{r_0^{3/2}}{\surd 2}\left\{\left[x(1-x)\right]^{1/2}+\arctan\left(\frac{1-x}{x}\right)^{1/2}\right\}
+\frac 13 r_0^3(1-x^3),
\label{eq:r0}
\eeqa
where $x\equiv r/r_0$.
The time at which the gas is accreted at the origin ($x=0$) is
\beq
t_a=\left(\frac{\pi}{2^{3/2}}\right)r_0^{3/2}+\frac 13 r_0^3\,.
\label{eq:ta}
\eeq
Note that this result is approximate, since it depends on the harmonic mean approximation
in equation (\ref{eq:ov}). We have found better agreement with the numerical results if
we approximate $t_a$ as the root mean square of the two terms in equation (\ref{eq:ta}):
\beq
t_a=\left(\frac{\pi^2 }{8}\, r_0^3+\frac 19\,  r_0^6\right)^{1/2}.
\eeq
The solution of this equation shows that gas accreting at time $t$ originated from
a radius $\roa$ given by
\beq
\roa^3=\left(\frac{9\pi^2}{16}\right)\frac{\tau^2}{1+(1+\tau^2)^{1/2}},
\label{eq:roa}
\eeq
where 
\beq
\tau\equiv \left(\frac{16}{3\pi^2}\right) t=0.540\,t.
\label{eq:tau}
\eeq
For late times ($t\gg1$), this reduces to
\beq
\roa\rightarrow (3t)^{1/3}.
\label{eq:latero}
\eeq

The accretion rate onto the origin is
\beq
\dot M=4\pi \lambda\roa^2\rho_\infty\,\frac{d\roa}{dt}~\times~\rb^2 c,
\eeq
where the final factor gives $\dot M$ the correct dimensions.
Evaluating the time derivative from equation (\ref{eq:roa}), we obtain
an approximation for the time-dependent accretion rate,
\beq
\dot M(t) \simeq 4\pi \rb^2\rho_\infty c\;\frac{\tau}{\left(1+\tau^2\right)^{1/2}}.
\label{eq:mdot}
\eeq
Thus, at early times the accretion rate increases linearly with time, whereas
at late times it approaches the steady state value given in equation (\ref{eq:dotms})
(although here we have set $\lambda=1$).

Consider now the particular case of steady flow.
Since the initial location of a mass element, $r_0$, depends on both $r$ and $t$, 
the steady flow approximation is valid only 
if the $r_0^{-1}$ term in equation (\ref{eq:ov}) is negligible. This is true
for $r\ll r_0$ or for sufficiently large $r_0$ provided $r$ is not too close to $r_0$.
As a check on the accuracy of equation (\ref{eq:ov}) in this case (i.e., when
$r_0^{-1}$ is negligible), 
note that the actual sonic point
is at $\rb/2$ \citep{ShapiroTeukolsky}, whereas equation (\ref{eq:ov}) gives $0.65\rb$;
the approximation is thus accurate to within about 30\%.
Equation (\ref{eq:steady}) gives the density for a steady flow, which requires that
the $r_0^{-1}$ term in equation (\ref{eq:ov}) be negligible:
\beq
\frac{\rho}{\rho_\infty}=\frac{1}{vr^2}\simeq 1+\frac{1}{\surd 2 r^{3/2}}
~~~~~\mbox{(steady flow)}.
\label{eq:rho}
\eeq

\subsection{The Magnetic Field}

When gas accretes onto the central object, both its mass and its pressure
are removed from the ambient medium. In the case of the magnetic field,
we assume that the flux is not accreted by the central star. As a result,
the flux associated with the accreted matter, $\Phi_a$, builds up and distorts
the flow close to the central object. When a flux tube loses
mass, it becomes buoyant and drives an interchange instability.
However, gas continues to accrete along
this flux tube so it may eventually fall back to the center. We therefore expect
that the innermost region will become turbulent. We begin with a discussion of
the magnetic field in the absence of the effects of the accretion flux, and then
estimate its effect at the end.

\subsubsection{The Field in Smooth Inflow}

Just as the gravitational force due to the star becomes important at radii
less than the Bondi radius, $\rb$, in the hydrodynamic case, so we expect
it to become important at radii less than the \alfven-Bondi radius,
\beqa
\rab&\equiv& \frac{GM_*}{\va^2}=\frac{4\pi GM_*\rho_\infty}{B_0^2},\label{eq:ra}\\
&=& 3.32\times 10^{15} \frac{M_*/M_\odot}{(\va/2\;\mbox{km s\e})^2}~~~~\mbox{cm}
\eeqa
in the MHD case. The ratio of the \alfven-Bondi radius to the standard
Bondi radius is
\beq
\frac{\rab}{\rb}=\frac{c^2}{\va^2}=\frac 12\, \beta,
\eeq
where $\beta\equiv 8\pi \rho_\infty c^2/B_0^2$ is the plasma $\beta$.
Our assumption that the field is weak implies $\beta\gg 1$.
There is an important relation between $\rab$ and
the magnetic critical mass
\beq
M_\Phi=\frac{\Phi}{2\pi G^{1/2}},
\eeq
which also determines the relative importance of self-gravity and magnetic fields:
\beq
\frac {\rab}{r_0}=\frac{4\pi G M_*\rho_\infty}{r_0B_0^2}=\frac 34\left(\frac{M_0 M_*}{M_\Phi^2}\right),
\eeq
where $M_0=4\pi\rho_\infty r_0^3/3$. The magnetic field is dominant for
$r_0>\rab$. In the purely gaseous case, the mass is subcritical for $M_0<M_\Phi$;
in the Bondi case, we see that the gas mass $M_0$ is replaced by the
geometric mean of the gas mass and the stellar mass (ignoring the
factor $\frac 34$). Shu, Li, \& Allen (2004) obtained a similar result for the case
in which the gas is in a disk; they
showed that it was possible for the field to be so strong that it could ``levitate" the
gas above a star in the process of formation.
Note that the fact that it is the geometric mean mass that determines whether
the gas is sub- or super-critical has an important consequence: in the purely
gaseous case, a sufficiently large uniform cloud is always supercritical,
since $M_0\propto r_0^3$ and $\Phi\propto r_0^2$. However, in the Bondi
case, the opposite occurs: a sufficiently large cloud is always subcritical,
since now $(M_0 M_*)^{1/2}\propto r_0^{3/2}$ increases more slowly than $\Phi$.

We assume that the field is initially uniform, so that $B_{\phi 0}=0$;
for spherical inflow, $B_\phi$ will remain zero.
For a spherical inflow, the radial flux through any surface $r^2d\Omega$ remains
constant, so that
\beq
r^2 B_r d\Omega =r_0^2 \bro d\Omega,
\eeq
which implies
\beq
B_r=\bro\left(\frac{r_0}{r}\right)^2=B_0\cos\theta \left(\frac{r_0}{r}\right)^2.
\label{A23}
\eeq

To evaluate $B_\theta$, consider a spherical shell of thickness $dr$ and
radius $r$. The flux in the shell at $\theta$ is proportional to $B_\theta r dr$.
The mass in the shell is $4\pi r^2\rho dr$. Since each of these remains
constant in the inflow, we have
\beq
r B_\theta dr \propto \rho r^2 dr,
\eeq
which implies
\beq
B_\theta=\bto\left(\frac{\rho r}{\rho_\infty r_0}\right)=
-B_0\sin\theta\left(\frac{\rho r}{\rho_\infty r_0}\right),
\label{A25}
\eeq
where the sign corresponds to the case in which the initial field is $\vecB_0=B_0\hat {\bf z}$.

How does the magnetic force compare with the gravitational one?  
First, we note that the
radial field by itself exerts no force; we therefore consider the pressure exerted by
$B_\theta$ and the tension force. We consider times late enough so that
$r_t\simeq (3t)^{1/3}$ and thus that $r_0$ is approximately independent
of $r$. For the pressure force, the relative importance of the magnetic field and
gravity in the midplane ($\theta=\pi/2$) can be assessed from the ratio
\beq
\frac{\va^2}{\vk^2}=\frac{B_\theta^2 r}{4\pi\rho GM}=\left(\frac{\rho}{\rho_\infty}\right)
\frac{r^3}{\rab r_0^2}.
\eeq
At large radii, we have $\rho\simeq\rho_\infty$; initially ($r\simeq r_0$) the
magnetic field dominates for $r>\rab$, as expected. At small radii,
$\rho/\rho_\infty\propto r^{-3/2}$ so that 
magnetic effects $\propto \va^2/\vk^2\propto r^{3/2}$ become negligible.

Next consider the tension in the radial direction, 
\beq
\frac{1}{4\pi}\;(\vecB\cdot\grad\vecB)_r=\frac{1}{4\pi}\left(\frac{B_\theta}{r}\,\ppbyp{B_r}{\theta}-\frac{B_\theta^2}{r}\right).
\eeq
The ratio of this force in the midplane to the gravitational force is
\beqa
\frac{F_{\rm tension}}{F_{g}}=\frac{(\vecB\cdot\grad\vecB)_r}{4\pi GM\rho/r^2}&=&
\frac{r_0}{\rab}\left(1-\frac{\rho r^3}{\rho_\infty r_0^3}\right)\\
&\rightarrow&\frac{r_0}{\rab}\left[1-\frac{r^3}{r_0^3}\left(1+\frac{1}{\surd 2 r^{3/2}}\right)\right],
\eeqa
where the last expression applies to steady flows. Provided $r_0\ga1$, the
density dependent term
becomes negligible for $r\ll r_0$, so that in this case
the force ratio becomes 
\beq
\frac{F_{\rm tension}}{F_{g}}\simeq \frac{r_0}{\rab}.
\eeq
Since $r_0\simeq (3t)^{1/3}$ at late times (eq. \ref{eq:latero}), it follows that the tension force
will eventually dominate and render the accretion anisotropic at
\beq
t_{\rm anis}\simeq \frac{\rab^3}{3c\rb^2}=\frac{\rb}{3c}\left(\frac{\beta}{2}\right)^3,
\eeq
where we have explicitly included the factors of $\rb$ and $c$. This is to be expected,
since as noted above a sufficiently large cloud is subcritical.

\subsubsection{Effects of the Accretion Flux} \label{accflux}

The accretion flux, $\Phi_a$, is the magnetic flux associated with the mass that
has accreted onto the central mass. We expect this flux to be buoyant and
to therefore lead to turbulence. Here we estimate the size of region affected
by the accretion flux.

At a time $t$, the accretion flux is the flux inside the initial
radius $\roa$ given in equation (\ref{eq:roa}), 
\beq
\frac{\Phi_a}{\pi\rb^2 B_0}= \roa^2 =
\left(\frac{9\pi^2}{16}\right)^{2/3}\frac{\tau^{4/3}}{\left[1+(1+\tau^2)^{1/2}\right]^{2/3}},
\eeq 
We estimate the radius, $r_\Phi$, out to which this flux extends
by assuming that the field associated with $\Phi_a$ is uniform, and
that the flow at $r_\Phi$ is steady. The latter assumption requires
that $r_\Phi$ be small compared to the starting radius, $r_0$, since
as discussed below equation (\ref{eq:ov}), $r_0(r, t)$ introduces time
dependent effects. We consider two limiting cases: (1) $r_\Phi\ll 1$,
where the accretion flux interacts with supersonic inflow and (2)
$r_\Phi>1$, where the accretion flux interacts with the pressure in
the ambient medium.

{\it Case 1: Supersonic inflow (early and intermediate times)}: We estimate $\rpo$, the
value of $r_\Phi$ in this case, by determining where the pressure due
to the accretion field balances the ram pressure of the accreting
gas. Since we are assuming that $\rpo\ll 1$ and $\rpo\ll r_0$,
equations (\ref{eq:ov}) and (\ref{eq:rho}) imply 
\beq
\frac{B_a^2}{8\pi}=\rho v^2=\frac{\surd 2\rho_\infty c^2}{\rpo^{5/2}} 
\eeq
Flux conservation implies $B_a \rpo^2=B_0 \roa^2$, so that 
\beqa
\rpo&=&\frac{\roa^{8/3}}{ 2^{1/3}
  \beta^{2/3}},\\ &=&\frac{1}{2^{1/3}\beta^{2/3}}
\left(\frac{9\pi^2}{16}\right)^{8/9} \frac{\tau^{16/9}}
     {\left[1+(1+\tau^2)^{1/2}\right]^{8/9}}\,.
\label{eq:rpo}
\eeqa
This expression is valid for both $\tau<1$ (early times) and $\tau>1$ (intermediate
times). At late times, the flow is dominated by thermal pressure.

{\it Case 2: Pressure-confined flow (late times)}: In this case the
magnetic pressure associated with the accretion flux balances the
thermal pressure of the ambient medium, 
\beq 
\frac{B_a^2}{8\pi}=\rho_\infty
c^2~~\Rightarrow~~\frac{B_a}{B_0}=\beta^{1/2}.  
\eeq 
Flux conservation then implies 
\beqa
\rpt&=&\frac{\roa}{\beta^{1/4}},\\ 
&=&\frac{1}{\beta^{1/4}}\left(\frac{9\pi^2}{16}\right)^{1/3}\frac{\tau^{2/3}}{\left[1+(1+\tau^2)^{1/2}\right]^{1/3}}.
\label{eq:rpt}
\eeqa

In order to obtain an approximation valid at all times, we write
\beq
\frac{1}{r_\Phi}\simeq \frac{1}{\rpo}\left(1+\frac{\rpo^2}{\rpt^2}\right)^{1/2}.
\label{eq:rphi}
\eeq 
Note that $r_\Phi$ is less than either $r_{\Phi,1}$ or ,
$r_{\Phi,2}$ corresponding to the fact that in this simple model the
pressure due to the escaped flux has to balance both the thermal
pressure and the ram pressure. Since $\rpo^2/\rpt^2$ exceeds unity
only at late times, this can be approximated as \beq r_\Phi\simeq
\frac{3.6
  \tau^{16/9}}{\beta^{2/3}\left[1+(1+\tau^2)^{1/2}\right]^{8/9}}
\cdot\frac{1}{\left(1+4.0\beta^{-5/6}\tau^{10/9}\right)^{1/2}}\, .
\eeq At early times, $r_\Phi\propto \tau^{16/9}$; at intermediate
times ($1\ll\tau\ll0.3\beta^{3/4}$), $r_\Phi\propto \tau^{8/9}$; and
at late times $r_\Phi\propto \tau^{1/3}$.

\section{Magnetic Bondi Flow in a Strong Magnetic Field}\label{B}

\subsection{Initial Transient}

A striking feature of Figure 2 for strong fields is that the flow is isotropic beyond
some radius, but then predominantly aligned along the axis inside that, until the flow is
very close to the center. This makes sense, since initially the field is straight and
therefore exerts no force; thus, at sufficiently early times, the flow for a strong
field is almost identical to that for no field. We focus on the region inside $\rb$, where
we neglect pressure forces. Let $r=r_0-\delta$, where $\delta\ll r_0$
since we are considering early times. Then equation (B5) implies
\beq
v=\frac{d\delta}{dt}=c\left(\frac{2\rb\delta}{r r_0}\right)^{1/2}\simeq c\,\frac{(2\rb\delta)^{1/2}}{r_0},
\eeq
where we have written the equation in dimensional form. Integration gives
\beq
\delta=\frac{\rb c^2t^2}{2 r_0^2}.
\eeq
The ratio of the tension force to the gravitational force at early times is given by
equation (B28) with $\rho=\rho_0$. For small $\delta$, this is
\beq
\frac{F_t}{F_g}=\frac{3\delta}{\rab}.
\eeq
The magnetic field will begin deflecting the flow from a radial trajectory to an
axial one when this ratio is of order unity, which occurs at
\beq
\frac{r_0}{\rb}=\left(\frac{3}{\beta}\right)^{1/2}\frac{t}{\tb}.
\eeq
We have found that the growth of the region deflected from a radial
trajectory in our numerical simulations with $\beta=0.1$ and $\beta=0.01$ follow this
functional form very well but that the deflection from spherical flow
occurs somewhat later than predicted.  We extract a good empirical fit
to the low-$\beta$ simulations with
\beq
\frac{r_0}{\rb}=\left(\frac{2.0}{\beta}\right)^{1/2}\frac{t}{\tb}.
\eeq

\subsection{Magnetic Bondi Flow in a Strong Magnetic Field ($\beta\la 0.1$) at Late Times} \label{B2}

For a very strong field, the gas will attempt to settle into vertical hydrostatic
equilibrium,
\beq
\rho=\rho_\infty e^{-m\phi/kT}=\rho_\infty e^{\rb/r}
\label{eq:rholowb}
\eeq
where $m$ is the mass per particle and $\phi=-GM_*/r$ is the gravitational potential.
Henceforth, we shall normalize all lengths to the Bondi radius, as in the previous section.
Outside the Bondi radius, this expression gives only a modest increase in density, but
for small radii the increase can be very large--so large that it takes a long time
to reach equilibrium. Let $\varpi$ be the cylindrical radius, so that $r=(\varpi^2+z^2)^{1/2}$,
where $z$ is the height above the disk. 
The density at the midplane ($r=\varpi$) is then
\beq
\rho_0=\rho_\infty e^{1/\varpi}.
\label{eq:rhoo}
\eeq
For small radii, $\varpi\ll1$, the
density distribution near the midplane is approximately gaussian,
\beq
\rho\simeq\rho_0 e^{-z^2/h^2},
\eeq
where $\rho_0$ is the midplane density and the scale height is
\beq
h=\surd 2 \varpi^{3/2}.
\eeq
In equilibrium, the total surface density of the gas near the midplane is then
\beq
\seq\simeq 2\rho_0 h=\rho_\infty\rb(2\varpi)^{3/2}e^{1/\varpi},
\label{eq:seq}
\eeq
where we have used equation (\ref{eq:rhoo}) to eliminate $\rho_0$.

When do magnetic forces balance gravity? For a thin disk, magnetic tension dominates
magnetic pressure \citep{shu97}. For an axisymmetric field,
the net radial tension is
\beq
F_t=\frac{1}{4\pi}(\vecB\cdot\vecnabla)B_\varpi= \frac{1}{4\pi\rb}B_z\ppbyp{B_\varpi}{z}.
\eeq
Integrating through the disk, we find
that the forces balance when
\beq
\frac{1}{4\pi}B_z(2B_\varpi)=\frac{GM_*\Sigma}{\rb\varpi^2},
\label{eq:balance}
\eeq
where $B_\varpi$ is measured just above the disk.

To obtain an accurate solution beyond this point, we would have to solve for the 
structure of the field. This is a challenging problem even when the system
is in equilibrium. Here, however, we are assuming that the system is in equililbrium
outside some critical radius, $\varcr$, but that there is an unknown accretion flow inside that
radius. We therefore content ourselves with attemping to infer the scaling for
the solution. We assume that $B_z$ in the disk is proportional to the ambient
field, $B_\infty$, and that the radial component of the field, $B_\varpi$, is proportional to 
$B_z$. Equation (\ref{eq:balance}) then implies that
\beq
\Sigma\sim \rho_\infty\rb\left(\frac{\varpi^2}{\beta}\right).
\label{eq:sigma}
\eeq
For a given location in the disk, gas will accrete along the field lines until the surface density
reaches this value. The field is unable to support more gas than this, so this value represents
an upper limit on $\Sigma$; any additional gas will accrete onto the central star.
However, we have determined another maximum value for the surface density
in equation (\ref{eq:seq}), which is the value the surface density has in hydrostatic
equilibrium. Equating these two surface densities determines the critical radius,
$\varcr$: The gas can be supported by the field outside $\varcr$, but inside $\varcr$ gas that
exceeds the surface density in equation (\ref{eq:sigma}) must fall onto the
central star. Equations (\ref{eq:seq}) and (\ref{eq:sigma}) imply that this critical radius
satisfies
\beq
\varcr^{1/2} e^{-1/\varcr}\sim \beta.
\eeq
A good approximation for the solution of this equation for $\beta\la 0.15$,
corresponding to $\varcr\la 0.6$, is
\beq
\varcr\simeq \frac{1}{\ln\beta^{-1}-0.5\ln\ln\beta^{-1}}~~~~~~~~(\beta\la 0.15).
\eeq
In the regime of greatest interest, $10^{-3}<\beta<0.15$, the solution can be approximated
by the simpler form
\beq
\varcr\simeq0.85\beta^{1/4}~~~~~~~~~~~~~(10^{-3}\la\beta\la 0.15).
\label{eq:approx2}
\eeq
The accuracy of this solution in the prescribed range is about 10\%, which is much
better than the accuracy of the underlying equation.

We are now in a position to estimate the accretion rate onto the central star. 
We assume that the accretion flow is primarily along the field lines, and that
it is initiated by a rarefaction wave propagating at the sound speed, $c$.
After an initial phase
during which the surface density just inside $\varcr$ becomes large enough
that it distorts the field so much that it can accrete, the accretion rate on
both sides of the disk becomes
\beq
\dot M\simeq 2 (\pi\rb^2\varocr^2)\rho_\infty c,
\eeq
where $\varocr$ is the cylindrical radius of the critical field lines far from the star.
If we assume that $\varcr\propto\varocr$, then in the range $10^{-3}\la\beta\la 0.1$
we have $\dot M\propto \varcr^2\propto \beta^{1/2}$, and we can write
\beq
\dot M=4\pi \llb\rb^2 \rho_\infty c\beta^{1/2},
\eeq
where $\llb$ is a numerical constant.  Note that the $\beta^{1/2}$
scaling is the same as that implied by the crude argument in the text.
Were we to assume that $\varocr=\varcr$ and that equation
(\ref{eq:approx2}) were accurate, then $\llb$ would equal 0.36. This
estimate is within a factor 1.6 of the numerical results. Setting
$\llb=0.24$ gives an accretion rate that agrees with the results of
the simulations for $\beta=0.1,\, 0.01$ to within 8\%.

\end{document}